
\input epsf                                                               %
\input harvmac
\input tables
\noblackbox
\def\Title#1#2{\rightline{#1}\ifx\answ\bigans\nopagenumbers\pageno0\vskip1in
\else\pageno1\vskip.8in\fi \centerline{\titlefont #2}\vskip .5in}

%
%
\ifx\epsfbox\UnDeFiNeD\message{(NO epsf.tex, FIGURES WILL BE IGNORED)}
\def\figin#1{\vskip2in}
\else\message{(FIGURES WILL BE INCLUDED)}\def\figin#1{#1}
\fi
\def\Fig#1{Fig.~\the\figno\xdef#1{Fig.~\the\figno}\global\advance\figno
 by1}
%
%
%
%
\def\ifig#1#2#3#4{
\goodbreak\midinsert
\figin{\centerline{\epsfysize=#4truein\epsfbox{#3}}}
\narrower\narrower\noindent{\footnotefont
{\bf #1:}  #2\par}
\endinsert
}
\def\ifigx#1#2#3#4{
\goodbreak\midinsert
\figin{\centerline{\epsfxsize=#4truein\epsfbox{#3}}}
\narrower\narrower\noindent{\footnotefont
{\bf #1:}  #2\par}
\endinsert
}
%
%
\def\calO{{\cal O}}
\def\tB{{\tilde B}}
\def\gtwid{\mathrel{\raise.3ex\hbox{$>$\kern-.75em\lower1ex\hbox{$\sim
$}}}}
\def\ltwid{\mathrel{\raise.3ex\hbox{$<$\kern-.75em\lower1ex\hbox{$\sim
$}}}}
\def\otp{{1\over2 \pi}}
\def\overleftrightarrow#1{\buildrel\leftrightarrow\over#1}

\def\sq{{\vbox {\hrule height 0.6pt\hbox{\vrule width 0.6pt\hskip 3pt
   \vbox{\vskip 6pt}\hskip 3pt \vrule width 0.6pt}\hrule height
0.6pt}}}
\def\vhat{{\widehat v}}

\def\bhat{{\widehat b}}

\def\ie{{\it i.e.}}
\def\invac{|0\rangle_{\rm in}}

\def\hf{{1\over2}}
\def\LDV{linear dilaton vacuum}

\font\ticp=cmcsc10

\def\aodag{a_\omega^\dagger}
\def\bodag{b_\omega^\dagger}
\def\bhodag{\bhat_\omega^\dagger}
\def\aopdag{a_{\omega'}^\dagger}

\def\sigmap{\sigma^+}
\def\sigmam{\sigma^m}

\def\ldv{linear dilaton vacuum}
\def\vbra{\langle 0|}
\def\vket{|0\rangle }
\def\sigmatm{{\tilde\sigma}^-}
\def\sigmahm{{\hat\sigma}^-}
\def\sigmahp{{\hat\sigma}^+}
\def\sigmah{{\hat \sigma}}
\def\sigmam{\sigma^-}
\def\sigmap{\sigma^+}
\def\xih{{\hat \xi}}
\def\xihm{{\hat\xi}^-}

\def\xim{\xi^-}

\def\scrip{{\cal I}^+}
\def\scrim{{\cal I}^-}

\def\Hsl{{\,\raise.15ex\hbox{/}\mkern-12mu H}}
\font\ticp=cmcsc10
\def\RN{Reissner-Nordstrom}

\def\sq{{\vbox {\hrule height 0.6pt\hbox{\vrule width 0.6pt\hskip 3pt
   \vbox{\vskip 6pt}\hskip 3pt \vrule width 0.6pt}\hrule height
0.6pt}}}
\def\jou#1&#2(#3){\unskip, \sl#1\bf#2\rm(19#3)}
\def\ajou#1&#2(#3){\ \sl#1\bf#2\rm(19#3)}
\def\frac#1#2{{#1 \over #2}}

\def\hf{{1\over2}}
\def\ie{{\it i.e.}}

\def\eg{{\it e.g.}}

\def\p+{{\partial_+}}

\def\delsl{\,\raise.15ex\hbox{/}\mkern-13.5mu \nabla}
\def\Ssl{{\,\raise.15ex\hbox{/}\mkern-10.5mu S}}

\def\tpl{t_{\rm pl}}
\def\mpl{m_{\rm pl}}
\def\rpl{r_{\rm pl}}

%
%
\lref\StrLH{A. Strominger, ``Les Houches lectures on black holes,'' to
appear.}
\lref\Thor{L. Thorlacius, ``Black hole evolution,'' hep-th/9411020,
NSF-ITP-94-109 (lectures at 1994 Trieste Spring School).}
\lref\Poly{A.M. Polyakov, ``Quantum geometry of bosonic strings,''\ajou
Phys. Lett. &103B (81) 207.}
\lref\CGHS{C.G. Callan, S.B. Giddings, J.A. Harvey and A. Strominger,
``Evanescent black holes,'' hep-th/9111056 \jou Phys. Rev. &D45 (92) R1005.}
\lref\RST{J.G. Russo, L. Susskind, and L. Thorlacius, ``Black hole
evaporation in 1+1 dimensions,'' hep-th/9201074\jou  Phys. Lett.
&B292 (92) 13.}
\lref\Schw{J. Schwinger, ``On gauge invariance and vacuum
polarization,''\ajou Phys. Rev. &82 (51) 664.}
\lref\AfMa{I.K. Affleck and N.S. Manton, ``Monopole pair production
in a
magnetic field,''\ajou Nucl. Phys. &B194 (82) 38.}
\lref\GaSt{D. Garfinkle and A. Strominger, ``Semiclassical Wheeler
wormhole
production,''\ajou Phys. Lett. &B256 (91) 146.}
\lref\deAl{S.P. deAlwis, ``Quantization of a theory of 2d dilaton
gravity,'' hep-th/9205069\jou Phys. Lett. &B289 (92) 278\semi
``Black hole physics from Liouville theory,'' hep-th/9206020\jou
Phys. Lett. &B300 (93) 330.}
\lref\Gibb{G.W. Gibbons, ``Quantized flux tubes in Einstein-Maxwell
theory
and  noncompact internal
spaces,'' in {\sl Fields and geometry}, proceedings of
22nd Karpacz Winter School of Theoretical Physics: Fields and
Geometry, Karpacz, Poland, Feb 17 - Mar 1, 1986, ed. A. Jadczyk
(World
Scientific, 1986).}
\lref\StTr{A. Strominger and S. Trivedi, ``Information consumption by
Reissner-Nordstrom black holes,'' hep-th/9302080
\jou Phys. Rev. & D48 (93) 5778.}
\lref\QTDG{S.B. Giddings and A. Strominger, ``Quantum theories of dilaton
gravity,'' hep-th/9207034\jou Phys. Rev. &D47 (93) 2454.}
\lref\PoSt{J. Polchinski and A. Strominger, ``A possible resolution of
 the black hole information puzzle,'' hep-th/9407008, UCSB-TH-94-20.}
\lref\StU{A. Strominger, ``Unitary rules for black hole evaporation,''
hep-th/9410187, UCSBTH-94-34.}
\lref\ItZu{C. Itzykson and J.-B. Zuber, {\sl Quantum Field Theory} (McGraw
Hill, 1980).}
\lref\StTh{A. Strominger and L. Thorlacius, ``Conformally invariant
boundary  conditions for dilaton gravity,'' hep-th/9405084\jou Phys. Rev.
&D50 (94) 5177.}
\lref\ChVe{T.D. Chung and H. Verlinde, ``Dynamical moving mirrors and black
holes,'' hep-th/9311007\jou Nucl. Phys. &B418 (94) 305.}
\lref\tHoo{G. 't Hooft, ``Fundamental aspects of quantum theory
related to
the problem of quantizing black holes,''  in {\sl Quantum Coherence},
ed.
J.S. Anandan (World Sci., 1990), and references therein\semi
Talk at 1993 ITP Conference, Quantum
Aspects of Black Holes.}
\lref\BaOl{T. Banks and M. O'Loughlin, ``Classical and quantum production
of cornucopions at energies below $10^{18}$ GeV,'' hep-th/9206055
\jou Phys.Rev. &D47 (93) 540-553.}
\lref\BOS{T. Banks, M. O'Loughlin, and A. Strominger, ``Black hole remnants
and the information puzzle,'' hep-th/9211030 \jou Phys. Rev.& D47 (93) 4476.}
\lref\Hux{T.H. Huxley, ``On a piece of chalk,'' in T.H. Huxley, {\sl
Essays} (Macmillan, NY, 1929).}
\lref\Wald{ R. M. Wald, ``On particle creation by black holes,''
\ajou Comm. Math. Phys. &45 (75) 9.}
\lref\MSW{G. Mandal, A Sengupta, and S. Wadia, ``Classical solutions of
two-dimensional
string theory,'' \ajou Mod. Phys. Lett. &A6 (91) 1685.}
\lref\Witt{E. Witten, ``On string theory and black holes,''\ajou Phys. Rev.
&D44 (91) 314.}
\lref\Erns{F. J. Ernst, ``Removal of the nodal singularity of the
C-metric,'' \ajou J. Math. Phys. &17 (76) 515.}
\lref\GiNe{S.B. Giddings and W.M. Nelson, ``Quantum emission from
two-dimensional black holes,''
hep-th/9204072\jou Phys. Rev. &D46 (92) 2486.}
\lref\BiDa{N.D. Birrell and P.C.W. Davies, {\sl Quantum fields in
curved
space} (Cambridge U.P., 1982).}
\lref\Japan{S.B. Giddings, ``Black holes and quantum predictability,''
hep-th/9306041, in {\sl Quantum
Gravity},  (Proceedings of the 7th
Nishinomiya Yukawa Memorial Symposium), K. Kikkawa and M. Ninomiya, eds.
(World Scientific, 1993).}
\lref\HaSt{J. Harvey and A. Strominger, ``Quantum aspects of black
holes,'', hep-th/9209055; in {\sl Recent Directions in Particle Theory,}
proceedings of the Theoretical Advanced Study
Institute, Boulder, CO, Jun 1992, J. Harvey and J. Polchinski, eds.
(World Scientific, 1993).}
\lref\HawkEvap{S.W. Hawking, ``Particle creation by black
holes,"\ajou Comm. Math. Phys. &43 (75) 199.}
\lref\HawkUnc{S.W. Hawking, ``The unpredictability of quantum
gravity,''\ajou Comm. Math. Phys &87 (82) 395.}
\lref\BPS{T. Banks, M.E. Peskin, and L. Susskind, ``Difficulties for
the
evolution of pure states into mixed states,''\ajou Nucl. Phys. &B244
(84)
125.}
\lref\Beke{J.D. Bekenstein, ``Black holes and entropy,''\ajou Phys. Rev. &
D7 (73) 2333.}
\lref\SVV{K. Schoutens, H. Verlinde, and  E. Verlinde, ``Quantum
black hole evaporation,''  hep-th/9304128\jou Phys. Rev. &D48 (93) 2670.}
\lref\HVer{H. Verlinde, lectures in this volume.}
\lref\PiSt{T. Piran and A. Strominger, ``Numerical analysis of black hole
evaporation,'' hep-th/9304148\jou Phys. Rev. &D48 (93) 4729.}
\lref\Lowe{D.A. Lowe, ``Semiclassical approach to black hole evaporation,''
hep-th/9209008\jou Phys. Rev. &D47 (93) 2446.}
\lref\Tada{T. Tada and S. Uehara, ``Consequence of Hawking radiation from
2-d dilaton black holes,'' hep-th/9409039, Kyoto Univ. preprint
YITP-U-94-23.}
\lref\CaWi{R.D. Carlitz and R.S. Willey, ``Reflections on moving
mirrors,''\ajou Phys. Rev. &D36 (87) 2327; ``Lifetime of a black
hole,''\ajou Phys. Rev. &D36 (87) 2336.}
\lref\Suss{L. Susskind, L. Thorlacius, and J. Uglum, ``The stretched horizon
and black hole complementarity,'' hep-th/9306069
\jou Phys. Rev. &D48 (93) 3743\semi
L. Susskind, ``String theory and the principles of black
hole
complementarity,'' hep-th/9307168\jou Phys. Rev. Lett. &71 (93) 2367\semi
L. Susskind and L. Thorlacius, ``Gedanken experiments involving black
holes,''  hep-th/9308100\jou Phys. Rev. &D49 (94) 966\semi
L. Susskind, ``Strings, black holes and Lorentz contraction,''
hep-th/930813\jou Phys. Rev. &D49 (94) 6606.}
\lref\Pres{J. Preskill, ``Do black holes destroy information?''
hep-th/9209058, in the
proceedings of the International Symposium on Black holes, Membranes,
Wormholes and Superstrings,
Woodlands, TX, 16-18 Jan 1992.}
\lref\Sred{M. Srednicki, ``Is purity eternal?,'' hep-th/920605 \jou
Nucl. Phys. &B410 (93) 143.}
\lref\Page{D. Page, ``Information in black hole radiation,''
hep-th/9306083\jou Phys. Rev. Lett. &71 (93) 3743.}
\lref\tHooaut{G. 't Hooft, ``Dimensional reduction in quantum gravity,''
gr-qc/9310006, Utrecht preprint THU-93/26, and references therein.}
\lref\LSU{D.A. Lowe, L. Susskind, and J. Uglum, ``Information spreading in
interacting string field theory,'' hep-th/9402136\jou Phys. Lett. &B327
(94) 226.}
\lref\CBHR{S.B. Giddings, ``Constraints on black hole remnants,''
hep-th/930402 \jou Phys. Rev. &D49 (94) 947.}
\lref\DGKT{H.F. Dowker, J.P. Gauntlett, D.A. Kastor and J. Traschen,
``Pair creation of dilaton black holes," hep-th/9309075
\jou Phys. Rev. & D49 (94) 2209.}
\lref\DGGH{F. Dowker, J. Gauntlett, S.B. Giddings, and G.T. Horowitz,
``On pair creation of extremal black holes and Kaluza-Klein
monopoles,'' hep-th/9312172 \jou Phys. Rev. &D50 (94) 2662.}
\lref\Melv{M. A. Melvin, ``Pure magnetic and electric
geons,''\ajou Phys. Lett. &8 (64) 65.}
\lref\GGS{D. Garfinkle, S.B. Giddings, and A. Strominger, ``Entropy in
Black Hole Pair Production,'' gr-qc/9306023 \jou Phys. Rev. &D49 (94) 958.}
\lref\BiCh{K. Bitar and S.-J. Chang, ``Vacuum tunneling and fluctuations
around a most probable escape path,''\ajou Phys. Rev. & D18 (78) 435.}
\lref\CILAR{S.B. Giddings, ``Comments on information loss and remnants,''
 hep-th/9310101 \jou Phys. Rev. & D49 (94) 4078.}
\lref\WABHIP{S.B. Giddings, ``Why aren't black holes infinitely
produced?'' (to appear).}
\lref\herring{S. Coleman, ``Black Holes as Red Herrings:
Topological Fluctuations and the Loss of Quantum Coherence,''\ajou Nucl.
Phys. & B307 (88) 864.}
\lref\LInc{S.B. Giddings and A. Strominger, ``Loss
of Incoherence and Determination of Coupling Constants
in Quantum Gravity,"\ajou Nucl. Phys. &B307 (88) 854.}
\lref\BiCa{A. Bilal and C. Callan, ``Liouville models of black hole
evaporation,'' hep-th/9205089\jou Nucl. Phys. &B394 (93) 73.}
\Title{\vbox{\baselineskip12pt\hbox{UCSBTH-94-52}\hbox{hep-th/9412138}
}}
{\vbox{\centerline {Quantum Mechanics of Black Holes}
}}
\centerline{{\ticp Steven B. Giddings}\footnote{$^\dagger$}
{Email addresses:
giddings@denali.physics.ucsb.edu}
}
\vskip.1in
\centerline{\sl Department of Physics}
\centerline{\sl University of California}
\centerline{\sl Santa Barbara, CA 93106-9530}

\bigskip
\centerline{\bf Abstract}
These lectures give a pedagogical review of
dilaton gravity, Hawking radiation,
the black hole information problem, and black hole pair creation.
(Lectures presented at the 1994 Trieste Summer School in High Energy Physics
and Cosmology)

\Date{}
\noblackbox

\newsec{Introduction}

Hawking's 1974 discovery\refs{\HawkEvap} that black holes evaporate
ushered in
a new era in black hole physics.  In particular, this was the
beginning of
concrete applications of quantum mechanics in the context of black
holes.
But more importantly, the discovery of Hawking evaporation has raised
a  sharp  problem whose resolution probably requires a better
understanding of Planck scale physics, and which therefore may serve
as
a guide (or at least a constraint) in our attempts to understand such
physics.  This problem is the information problem.

In brief, the information problem arises when one considers a
Gedanken
experiment where a  black hole is formed in collapse of a carefully
arranged
{\it pure} quantum state $\vert \psi\rangle $,
or in terms of quantum-mechanical
density matrices, $\rho=\vert\psi\rangle \langle \psi\vert$.  This
black
hole then evaporates, and according to Hawking's calculation the
resulting
outgoing state is approximately thermal, and in particular is
a {\it
mixed}
quantum state.  The latter statement means that the density matrix is
of
the form $\sum_\alpha p_\alpha \vert \psi_\alpha\rangle \langle
\psi_\alpha \vert$, for some normalized
basis states $\vert \psi_\alpha\rangle $ and some real
numbers $p_\alpha$ more than one of which is non-zero.
In the present case the $p_\alpha$ are approximately
the usual thermal probabilities.  If Hawking's
calculation can be trusted, and if the black hole does not leave behind a
remnant,
this means that in the quantum theory of
black
holes pure states can evolve to mixed.  This conflicts with the
ordinary
laws of quantum mechanics, which always preserve purity.
Comparing pure and
mixed states, we find that there is missing phase information in the
latter.  A useful
measure of the missing information is the entropy, $S=-{\rm Tr}
\rho \ln \rho =- \sum_\alpha p_\alpha \ln p_\alpha$.

Hawking subsequently proposed \refs{\HawkUnc} that quantum mechanics
be
modified to allow purity loss.  However, as we'll see, inventing an
alternative dynamics is problematical.  This has lead people to
consider
other alternatives, namely that information either escapes the black
hole or
that it is left behind in a black hole remnant.  Both of these
possibilities also encounter difficulties, and as a result we have
the
black hole information problem.

These lectures will develop these statements more fully.\foot{Other
reviews include
\refs{\HaSt,\Japan} and more recently \refs{\Thor,\StrLH}.}
In the past few years a much advanced
understanding of black hole evaporation has been obtained through
investigation of
two-dimensional models, and because of this and due to their greater
simplicity we'll start by reviewing these models.  Next
will be a detailed treatment of Hawking radiation from the resulting
two-dimensional black holes, followed by a generalization to the derivation
of four-dimensional Hawking radiation.  As black holes Hawking radiate they
shrink, and a subsequent section is devoted to a semiclassical description
of such black hole evaporation.  There follows a review of the black hole
information problem, its various proposed resolutions, and the problems
with these proposals.  Finally we turn to a treatment of another
non-trivial aspect of the quantum mechanics of black holes, black hole pair
production.  Besides its intrinsic interest, this process can potentially
shed light on the proposed
remnant resolution of the information problem, and a
brief discussion of how it does so is given.

\newsec{Two-dimensional dilaton gravity}

In 2d, formulating gravity with just a metric gives trivial dynamics;
for
example, the Einstein action is a topological invariant.  Instead we
consider theories with the addition of a scalar dilaton $\phi$.  A
particular simple theory\refs{\CGHS} has action
\eqn\sgrav{ S= { 1 \over 2\pi}\int d^2
x\sqrt{-g}\left[e^{-2\phi}(R+4(\nabla\phi)^2
+4\lambda^2)-
\half(\nabla f)^2\right]\ ,}
where $\lambda^2$ is an analogue to the cosmological constant and $f$
is a
minimally coupled massless matter field that provides a source for
gravity.  Note that $e^\phi$ plays the role of the gravitational
coupling,
as its inverse square appears in front of the gravitational part
of the action.

In two dimensions the general metric $ds^2=g_{ab}dx^adx^b$ can always
locally be put into conformal gauge,
\eqn\confg{ds^2 = -e^{2\rho} dx^+dx^-\ ,}
with the convention  $x^\pm = x^0\pm x^1$.  The equations
resulting from the action \sgrav\ are most easily analyzed in this
gauge.
The matter equations are
\eqn\matteq{\partial_+\partial_- f=0\ ,}
with general solution $f_i= f_+(x^+) + f_-(x^-)$.  Next, the relation
\eqn\curva{{\sqrt -g} R = -2 \sq \rho\ }
allows rewriting of the gravitational part of the action,
\eqn\cgact{ S_{\rm grav} = {1\over 2 \pi} \int d^2 x \left\{ 2
\nabla(\rho-\phi)
\nabla e^{-2\phi} + 4\lambda^2 e^{2(\rho-\phi)}\right\}\ .}
The equation of motion for $\rho-\phi$ is therefore that of a free
field,
with solution
\eqn\rhophi{\rho-\phi =\hf\left[w_+(x^+) + w_-(x^-)\right]\ . }
The equation for $\phi$ then easily gives
\eqn\phieq{e^{-2\phi} = u_+ + u_- - \lambda^2 \int^{x^+} dx^+
e^{w_+}\int^{x^-} dx^-
e^{w_-}\ }
where $u_\pm(x^\pm)$ are also free fields.  Finally, varying the
action \sgrav\
with respect to $g^{++}$, $g^{--}$ gives the constraint
equations,
\eqn\consEOM{\eqalign{
\delta g^{++}:\ &\ G_{++}\equiv-e^{-2\phi}\left(  4\partial_+\rho
\partial_+\phi-2\partial^2_+ \phi\right) =\half\partial_+ f
\partial_+f\cr
\delta g^{--}:\ &\
G_{--}\equiv-e^{-2\phi}\left(4\partial_-\rho\partial_-\phi-2\partial^2
_
-\phi\right)
 =\half\partial_-f\partial_-f\ .}}
These determine $u_\pm$ in terms of $f_\pm$ and
$w_\pm$:
\eqn\ueq{u_\pm = {M\over2\lambda }  -\half \int dx^\pm e^{w_\pm}
\int dx^\pm e^{-w_\pm} \partial_\pm
f\partial_\pm f\ ,}
where $M$ is an integration constant.  In the following we will
choose
units so that $\lambda=1$.

The theory is therefore completely soluble at the classical level.
The unspecified functions $w_\pm$ result from the unfixed remaining
gauge
freedom; conformal gauge \confg\ is unchanged by a reparametrization
$x^\pm
=x^\pm(\sigma^\pm)$.  This freedom may be used to set $w_++w_-
=\sigma^+-\sigma^-$,
for example.
In
this gauge the general vacuum solution is\refs{\MSW,\Witt}
\eqn\gsoln{
\eqalign{
ds^2 & = -\frac{d\sigma^+ d\sigma^-}{1+Me^{\sigma^--\sigma^+}}\cr
\phi & = -\half \ell n \left( M+e^{\sigma^+-\sigma^-}\right)\ .\cr
}}
The case $M=0$ corresponds to the ground state,
\eqn\LDV{\eqalign{
ds^2 & = -d\sigma^+ d\sigma^-\cr
\phi & = -\sigma\ .\cr
}}
This is the dilaton-gravity analogue of flat Minkowski space, and is
called the
linear dilaton vacuum.

The solutions for $M>0$ are asymptotically flat as
$\sigma^+ -\sigma^-\rightarrow \infty$.  At $\sigma^+
-\sigma^-\rightarrow
-\infty$ they are apparently singular, but regularity is restored by
the
coordinate transformation
\eqn\coordx{x^+ = e^{\sigma^+}\ , x^- = -e^{-\sigma^-}\ .}
A true curvature singularity appears at $x^+x^-=M$, and $x^\pm=0$ is the
horizon.
The corresponding Penrose
diagram is shown in Fig.~1; the solution is a black hole and $M$ is
its
mass.  Notice the important relation $e^{2\phi}|_{\rm
horizon}=\frac{1}{M}$.  For $M<0$ the solution is a naked
singularity.

\ifig{\Fig\Vpen}{Shown is the Penrose diagram for a vacuum
two-dimensional dilatonic black hole.}{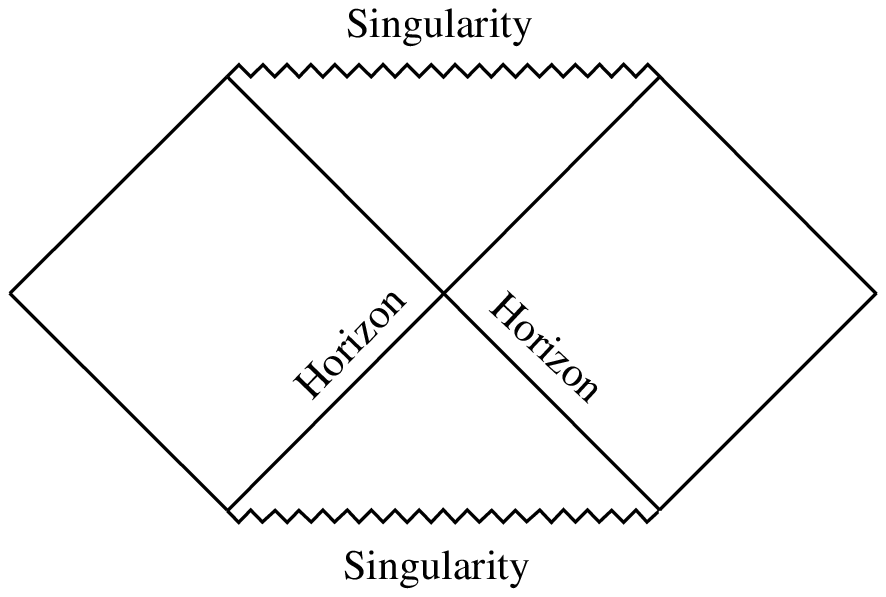}{1.75}

\ifig{\Fig\cpen}{The Penrose diagram for a collapsing black hole
formed from a left-moving matter distribution.}{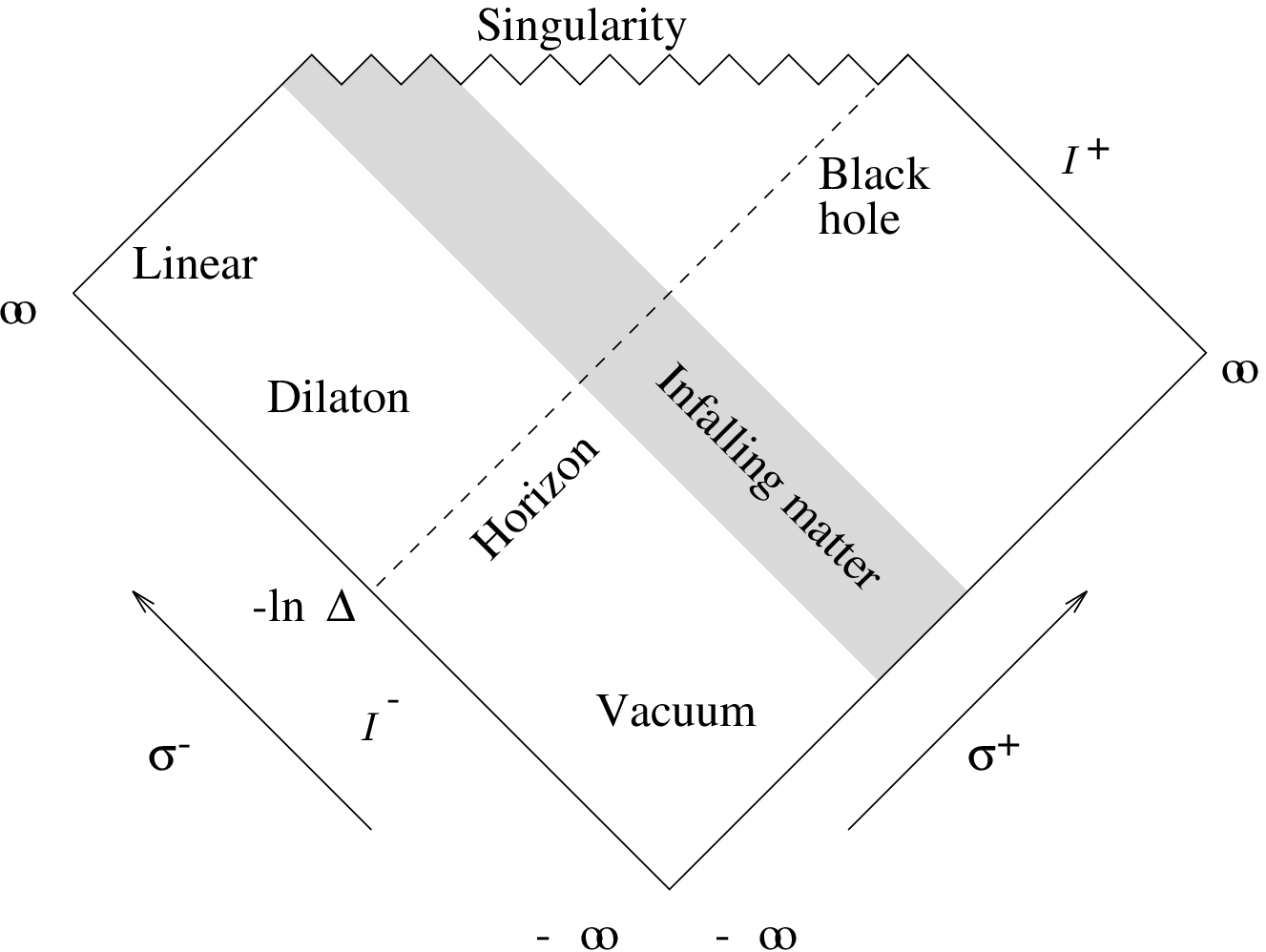}{3.00}

Next consider sending infalling matter, $f=F(x^+)$, into the \ldv.
This will form a black hole, as shown in Fig.~2.  Before the matter
infall the solution is given by \LDV. Afterwards it is found by using
\rhophi-\ueq,
\eqn\fmet{\eqalign{ e^{-2\phi}&= M+
e^{\sigma^+}\left(e^{-\sigma^-}-\Delta\right)\cr
ds^2 &= -\frac{d\sigma^+ d\sigma^-}{1+M\,e^{\sigma^- - \sigma^+} -
\Delta
\, e^{\sigma^-}}}
}
where one can easily show
\eqn\six{
\eqalign{
M = & \int d\sigma^+ T_{++}\cr
\Delta = & \int d\sigma^+ e^{-\sigma^+}T_{++}\ ,\cr
}}
and where
\eqn\seven{
T_{++}=\half(\partial_+ F)^2
}
is the stress tensor.  The coordinate transformation
\eqn\eight{
\xi^- = -\ell n\left(e^{-\sigma^-} - \Delta\right)\ ,\ \xi^+ =
\sigma^+
}
returns the metric \fmet\ to the asymptotically flat form
\eqn\nine{
ds^2 = -\frac{d\xi^+ d\xi^-}{1+M\ e^{\xi^- - \xi^+}}\ ,
}
showing that indeed we have formed a black hole.

\newsec{Hawking radiation in two dimensions}

With a collapsing black hole in hand, we can study its Hawking
radiation.\foot{For other references see
\refs{\HawkEvap,\BiDa,\GiNe}.}
The quickest derivation of the Hawking flux arises by
computing the expectation value of the matter stress tensor.
Consider the stress tensor for right-movers; before the hole forms
they are
in their vacuum, and
\eqn\Texpec{\eqalign { \langle T_{--} \rangle &=
\lim_{\sigmahm\rightarrow
\sigma^-} \langle 0| \hf\partial_- f(\sigmahm) \partial_- f(\sigma^-)
|0\rangle \cr
&=\lim_{\sigmahm\rightarrow
\sigma^-}  {-1\over 4(\sigmahm-\sigma^-)^2} \ ,}}
where the second line uses the 2d Green function,
\eqn\tdgr{ \vbra f(\sigmah) f(\sigma)\vket =
-\hf\left[\ln(\sigmahp -\sigmap)+\ln(\sigmahm-\sigmam) \right]\ .}
As usual, one removes the infinite vacuum energy by normal-ordering:
\eqn\nT{:T_{--}:_\sigma = T_{--} +{1\over 4}
{1\over(\sigmahm-\sigma^-)^2} \ .}
The formula \Texpec\ can also be applied at $\scrip$, but the flat
coordinates are now $\xi^\pm$.  Therefore to compare the stress
tensor to
that of the outgoing vacuum on $\scrip$, we should subtract the
vacuum energy
computed in the $\xi$ coordinates,
\eqn\nxT{:T_{--}:_\xi = T_{--} + {1\over4(\xihm-\xi^-)^2} \ .}
The corresponding expectation value is
\eqn\nnT{\eqalign{
\langle :T_{--}:_\xi \rangle &=\lim_{\xihm
\rightarrow \xim} \vbra \hf
\partial^\xi_- f\left(\sigma^-(\xihm)\right)
\partial^\xi_- f\left(\sigmam(\xim)\right)\vket +
{1\over 4(\xihm-\xim)^2}\cr
&= -{1\over 4}\lim_{\xihm\rightarrow
\xim} \partial_-^{\xih}\partial_-^\xi \ln
\left(\sigma^-(\xihm) -\sigma^-(\xim)\right) +{1\over4
(\xihm-\xim)^2}\ .}}
Next one expands $\sigma^-(\xihm)$ about $\xim$, and in a few lines
finds
\eqn\Tmm{\langle :T_{--}:_\xi\rangle = -{1\over 24} \left[
{\sigma^{-'''}\over \sigma^{-'}} - {3\over 2}
\left({\sigma^{-''}\over
\sigma^{-'}} \right)^2\right]\ }
where prime denotes derivative with respect to $\xim$.

Using the relation \eight\ between the two coordinate systems gives
the
outgoing stress tensor from the black hole,
\eqn\hawkst{ \langle :T_{--}:_\xi\rangle  =
\frac{1}{48} \left[1-\frac{1}{(1+\Delta e^{\xi^-})^2}\right]\ .
}
This exhibits transitory behavior on the scale $\xi^-\sim
-\ln\Delta$, but as
$\xi^-\rightarrow \infty$ it asymptotes to a constant value $1/48$.
As
will be seen shortly, this corresponds to the  thermal Hawking
flux at a temperature $T=1/2\pi$.

A more detailed understanding of the Hawking radiation arises from
quantizing the scalar field.  Recall the basic steps of canonical
quantization:

\item {1.}  Find a complete orthonormal
basis of solutions to the field equations.

\item {2.}  Separate these solutions into orthogonal sets of
positive and  negative frequency.

\item {3.} Expand the general field in terms of the basis functions
with annihilation operators as coefficients of positive frequency
solutions
and
creation operators for negative frequency.

\item {4.} Use the canonical commutation relations to determine the
commutators of these ladder operators.

\item {5.}  Define the vacuum as the state annihilated by the
annihilation
operators, and build the other states by acting on it with creation
operators.

In curved spacetime general coordinate invariance implies that step
two is
ambiguous:  what is positive frequency in one frame is not in
another.
Consequently the vacuum state is not uniquely defined.  These two
observations are at the heart of the description of particle creation
in
curved spacetime.  This ambiguity was responsible for the different
normal-ordering prescriptions in our preceding derivation.

Now follow this recipe. Begin by noting that
the equations of motion imply existence of
the conserved
Klein-Gordon inner product,
\eqn\kgprod{ (f,g)=-i\int_{\Sigma}{d\Sigma^{\mu}f
 {\overleftrightarrow\nabla}_{\mu} g^*}}
for arbitrary Cauchy surface $\Sigma$.

Steps 1 and 2:  A convenient basis for right-moving modes
in the ``in'' region near $\scrim$ are
\eqn\pnmodes{
u_\omega =  {1\over\sqrt{2\omega}}e^{-i\omega\sigma^-}\ ,\
u_\omega^* =  {1\over\sqrt{2\omega}}e^{i\omega\sigma^-}\ .
}
These have been normalized so that
\eqn\normcond{ (u_{\omega},u_{\omega^{\prime}})=
2\pi\delta(\omega-\omega^{\prime}) =
-({u}_{\omega}^*,{u}_{\omega^{\prime}}^*)\ ,\
(u_{\omega},{u}_{\omega^{\prime}}^*)=0\ .}
Furthermore, note that they are naturally separated according to
positive
or negative frequency with respect to the
time variable $\sigma^0$.

Step 3:  The field $f$ has expansion in terms of annihilation and
creation
operators
\eqn\fexp{ f_- = \int_0^\infty d\omega \left[a_{\omega}u_{\omega}
                        +\aodag {u}_{\omega}^*\right]\ ({\rm in})\ .}

Step 4:  The canonical commutation relations
\eqn\ccrp{ [f_-(x),\partial_{0}f_-(x^{\prime})]_{x^{0}=x^{\prime0} }
=\hf [f(x),\partial_{0}f(x^{\prime})]_{x^{0}=x^{\prime0}} = \pi i
\delta(x^1
- x^{\prime 1})}
together with the inner products \normcond\
imply that the operators $a_{\omega}$
satisfy the usual commutators,
\eqn\ccr{
[a_{\omega},\aopdag ]
= \delta(\omega - \omega^{\prime})\ ,\
 [a_{\omega},a_{\omega^{\prime}}]=0\ ,\
[\aodag,\aopdag]=0\ .}

Step 5:  The in vacuum is defined by
\eqn\invac{a_\omega\vket_{in} =0\ ;}
other states are built on it by acting with the $a_\omega^\dagger$'s.

To describe states in the future regions it is convenient to follow these
steps
with a different set of modes.  Modes are needed both in
the ``out'' region near $\scrip$ and
near the singularity.  The former are the obvious analogues to
\pnmodes,
\eqn\outmodes{v_\omega =  {1\over\sqrt{2\omega}}e^{-i\omega\xi^-}\ ,\
v_\omega^* =  {1\over\sqrt{2\omega}}e^{i\omega\xi^-}\ .}
The latter are somewhat arbitrary as the region near the singularity
is
highly curved.  A convenient coordinate near the singularity proves
to be
\eqn\csing{\xihm = \ln( \Delta^2 e^{\sigma^-}-\Delta)\ ,}
and corresponding modes $\vhat_\omega$ and $\vhat_\omega^*$ are given
by a
formula just like \outmodes.

In terms of these modes, $f$ is written
\eqn\fexpo{f_- = \int_0^\infty d\omega \left[b_\omega v_\omega +
\bodag
v_\omega^* + \bhat_\omega \vhat_\omega + \bhodag
\vhat_\omega^*\right] \ ({\rm out}+{\rm internal})\ .}
These modes are normalized as in \normcond, and the corresponding
field
operators obey commutators as in \ccr.  The vacua $\vket_{\rm out}$
and
$\vket_{\rm internal}$ are also defined
analogously to \invac.

The non-trivial relation \eight\
between the natural timelike coordinates in the
in and out regions -- or equivalently
between the symmetry directions with respect to which positive
and negative frequencies are defined --
imply that a positive frequency solution in one
region
is a mixture of positive and negative frequency in another region.
This
mixing implies particle creation.  For example, positive frequency
out
modes can be expressed in terms of the in modes,
\eqn\oimodes{
 v_{\omega} =\int_{0}^{\infty}d\omega' \lbrack
  \alpha_{\omega\omega^{'}} u_{\omega^{'}}
 +\beta_{\omega\omega^{'}} u^{*}_{\omega^{'}} \rbrack\ . }
The Fourier coefficients $\alpha_{\omega\omega'}$,
$\beta_{\omega\omega'}$ are called Bogoliubov coefficients, and they
can
be calculated by inverting the Fourier transform, or equivalently from
\eqn\bogc{\alpha_{\omega\omega^{\prime}} =\otp (v_{\omega},
                         u_{\omega^{\prime}})\quad ,\quad
\beta_{\omega\omega^\prime} = -\otp(v_{\omega},
                         {u}_{\omega^{\prime}}^*)\ .}
In the present model they can easily be found in closed form in terms
of
incomplete beta functions\refs{\GiNe}
\eqn\bogxm{\eqalign{ \alpha_{\omega\omega^{\prime}} &=
   {1\over2\pi}\sqrt{{\omega^{\prime}}\over{\omega}}
  {\Delta}^{ i\omega}
B\left(-{ i\omega}+{ i\omega^{\prime}}
   ,1+{ i\omega} \right) \cr
\beta_{\omega\omega^{\prime}}  &=
{1\over2\pi}\sqrt{{\omega^{\prime}}\over{\omega}}
\Delta^{ i\omega}
B\left(-{ i\omega}-{ i\omega^{\prime}}
,1+{ i\omega} \right)\ .}}
Although we will not use these formulas directly they are exhibited
for
completeness.

To investigate the thermal behavior at late times, $\xi^- \gg -\ln
\Delta$,
$\sigma^-\simeq-\ln\Delta$,
we could examine the asymptotic behavior of \bogxm, but a shortcut is
to
use the asymptotic form of \eight\ found by expanding around $\sigma^- = -
\ln \Delta$,
\eqn\coordr{ -e^{-\xim}= \Delta-e^{-\sigmam} \simeq \Delta(\sigma^-
+\ln\Delta)\equiv \sigmatm\ .}
Note that this is the same as the relation between Rindler and
Minkowski
coordinates in the context of accelerated motion.  Likewise one finds
\eqn\coordh{ e^{\xihm}\simeq \sigmatm\ .}
Next, a trick\refs{\Wald} can be used to
find the approximate form for the outgoing state.
Notice that functions that are positive frequency in $\sigmatm$
are analytic in the lower half complex $\sigmatm$ plane.
Therefore the functions
\eqn\modcomb{\eqalign{u_{1,\omega}\propto(-\sigmatm)^{i\omega}&\propto
v_\omega +e^{-\pi\omega}
\vhat_\omega^*\cr
u_{2,\omega}\propto(\sigmatm)^{-i\omega}&\propto \vhat_\omega
+e^{-\pi\omega}
v_\omega^* }}
are positive frequency.  This means that the
corresponding field operators $a_{1,\omega}$ and $a_{2,\omega}$ must
annihilate the in vacuum.  The inverse of the transformation
\modcomb\
gives the relation between field operators, as can easily be seen by
reexpanding \fexpo\ in the $u_{i,\omega}$'s and $a_{i\omega}$'s:
\eqn\invbog{\eqalign{
a_{1\omega} &\propto
b_\omega - e^{-\pi\omega} {\hat b}^\dagger_\omega\cr
a_{2\omega} &\propto {\hat b}_\omega - e^{-\pi\omega}
b^\dagger_\omega\ .}}

These then determine the vacuum, since it obeys
\eqn\killvac{\eqalign{0 &= (a_{1,\omega}^\dagger a_{1,\omega} -
a_{2,\omega}^\dagger
a_{2,\omega}) \vket \cr
&\propto (b_\omega^\dagger b_\omega - {\hat b}_\omega^\dagger {\hat
b}_\omega)
 \vket \cr
&\propto (N_\omega- {\hat N}_\omega)\vket\ ,}}
where $N_\omega$, ${\hat N}_\omega$ are the number operators for the
respective modes.  The latter equation implies that
\eqn\veqn{\vket = \sum_{\{n_\omega\}} c\left(\{n_\omega\}\right)
\widehat{|\{n_\omega\}\rangle } |\{n_\omega\}\rangle\ }
for some numbers $c\left(\{n_\omega\}\right)$.  These numbers are
determined up
to an overall constant from the recursion relation following from the
equation $a_{1\omega} \vket =0$:
\eqn\ceqn{c\left(\{n_\omega\}\right) = c\left(\{0\}\right)
\exp\left\{ -\pi\int
d\omega \omega n_\omega \right\}\ .}
Thus the state takes the form
\eqn\vacst{ \vket = c\left(\{0\}\right) \sum_{ \{n_\omega\} }
e^{-\pi\int
d\omega \omega n_\omega} \widehat {|\{n_\omega\}\rangle }
|\{n_\omega\}\rangle\ \ .}

It is clear from this relation that the state inside the black hole
is
strongly correlated with the state outside the black hole.  Observers
outside the hole cannot measure the state inside, and so summarize
their
experiments by the density matrix obtained by tracing over all
possible
internal states,
\eqn\dmat{ \rho^{\rm out} = {\rm Tr}_{\rm inside} \vket \vbra =
|c\left(\{0\}\right)|^2 \sum_{\{n_\omega\}} e^{-2\pi\int
d\omega \omega n_\omega} |\{n_\omega\}\rangle \langle \{n_\omega\}|\
.}
This is an exactly thermal density matrix with temperature
$T=1/2\pi$.
The corresponding energy density is that of a right moving scalar field,
\eqn\edens{{\cal E}= \int _0^\infty {d\omega\over 2\pi} {\omega\over
e^{\omega/T }-1} = {\pi\over 12}T^2= {1\over 48\pi}\ ,}
which agrees with \hawkst\ if we account for the
unconventional normalization
of the stress tensor used in \CGHS,
\eqn\Tnorm{ {\cal E} = {1\over \pi} T_{00}\ .}
Both the total entropy and energy of this density matrix are
infinite,
but that is simply
because we have not yet included backreaction which causes the
black hole to shrink
as it evaporates.   Before considering these issues, however, we will
extend these results to four-dimensional black holes.

\newsec{Hawking radiation in four dimensions}

\ifig{\Fig\fdpen}{The Penrose diagram for a collapsing
four-dimensional black hole.  Also indicated are lines of constant
$\sigma^-$ (solid) and lines of constant $\xi^-$ (dashed), as well as the
ray ${\cal R}$ described in the text.}{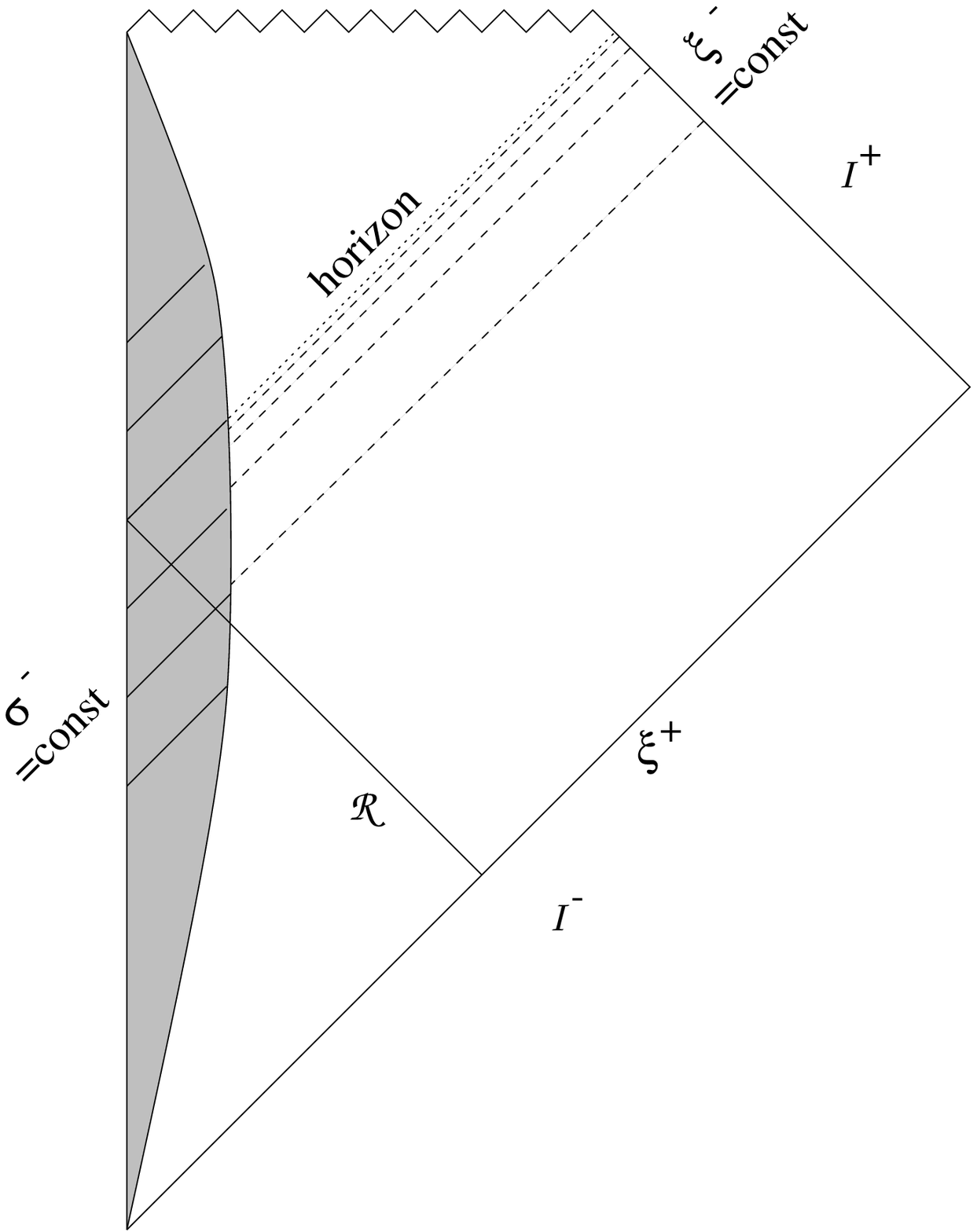}{3.5}

\fdpen\ shows the Penrose diagram for a collapsing four-dimensional
black hole; this picture is analogous to fig.~2. Making the
assumption of radial
symmetry implies the metric takes the form
\eqn\fdmet{ds^2=g_{tt} dt^2 + g_{rr} dr^2 + R^2 (r,t) d\Omega^2_2}
where $g_{tt}, g_{rr}$, and $R$ are functions of $r$ and $t$ and
$d\Omega^2_2$ is the line element on the two-sphere.  Notice that as
before reparametrizations of $r$ and $t$ allow us to rewrite \fdmet\
in
``conformal gauge'',
\eqn\fourtwo{ds^2 = e^{2\rho} \left[-(dx^0)^2 + (dx^1)^2\right] + R^2
(x) d\Omega^2_2\ .}
In particular, outside the black hole the metric is time independent
by
Birkhoff's theorem, and can be written
\eqn\outmet{ds^2 = -g_{tt} \left(-dt^2 + dr^{*2}\right) + r^2
d\Omega^2_2}
where the tortoise coordinate $r^*$ is defined by
\eqn\fourfour{\frac{dr^*}{dr} = \sqrt{\frac{g_{rr}}{-g_{tt}}}\ .}
(For the Schwarzschild black hole $r^* = r+ 2M \ell n (r-2M)$.)

Now consider propagation of a free scalar field,
\eqn\mincoup{S=-\half \int d^4 x \sqrt{-g} (\nabla f)^2\ .}
The solutions take the separated form
\eqn\foursix{f=\frac{u(r,t)}{R}\ Y_{\ell m} (\theta, \phi)}
in terms of spherical harmonics and two-dimensional solutions
$u(r,t)$.
To compare with our two-dimensional calculation, first rewrite
\mincoup\ in two-dimensional form.  For example, in the outside
coordinates \outmet,
\eqn\fourseven{\int d^4 x\sqrt{-g} (\nabla f)^2 \propto \int dr^* dt
\left[(\partial_t u)^2 - (\partial_{r^*} u)^2 - V(r^*) u^2\right]}
where the effective potential is
\eqn\foureight{V(r^*) = -g_{tt}\frac{\ell(\ell+1)}{r^2} +
\frac{\partial^2_{r^*}r}{r}\ .}
The latter term vanishes at $r=r^*=\infty$ as $\frac{1}{r^3}$, and
both
terms vanish as $r^*\to -\infty$.  For Schwarzschild
the maximum of the potential is
of order $\ell(\ell + 1)/M^2$, and
occurs where $r^*\sim M$.  These features are sketched in fig.~4.

\ifig{\Fig\effpot}{A sketch of the effective potential as a function of
tortoise coordinate $r^*$.  (Shown is the case of the Schwarzschild black
hole.)}{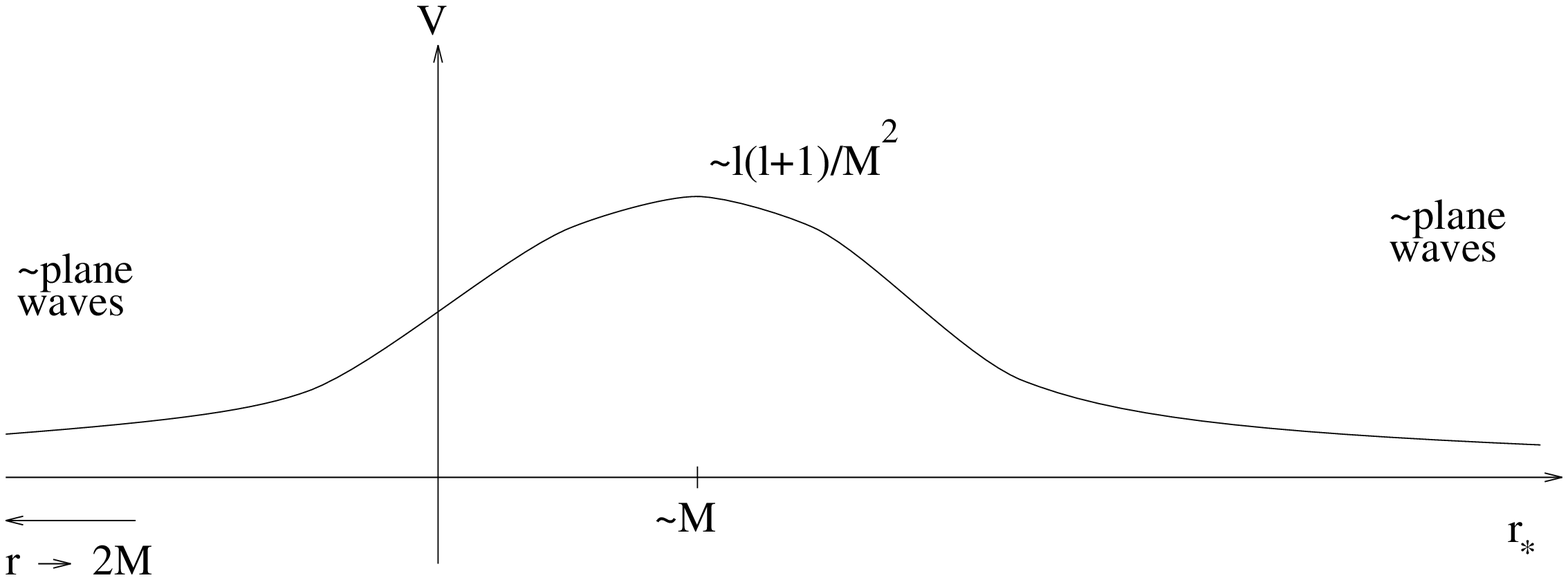}{1.5}

In the limits $r^*\to \pm \infty$ the solutions are therefore plane
waves.  However, an outgoing plane wave from $r^*=-\infty$ is only
partially transmitted.  As we'll see, the Hawking radiation can be
thought of as thermally populating the outgoing modes, and these
energy-dependent ``gray body'' transmission
factors lead to deviations from a pure
black body spectrum.  These factors are, however, negligible for
$\omega^2 \gg V_{\rm max}$, where for Schwarzschild
$V_{\rm max}\sim \ell(\ell+1)/M^2$.

The Hawking effect is again found by relating the in modes at ${\cal I}^-$
to the out modes at ${\cal I}^+$.
The
good asymptotically flat coordinates outside the collapsing body were
given in \outmet:
\eqn\fournine{ds^2 = -g_{tt}\ d\xi^+ d\xi^- + r^2 d\Omega^2_2 \quad,
\quad
\xi^\pm = t\pm r^*\ .}
These coordinates are suitable for the in or out regions. We will
also
use conformal coordinates for the interior of the collapsing body,
\eqn\fourten{ds^2 = -\Omega^2(\sigma)\ d\sigma^+ d\sigma^- +
r^2 d\Omega^2_2\ .}
In particular, we can label the slices so that
near the point where the surface of the collapsing
matter crosses the horizon
\eqn\fourtwelve{\sigma^\pm = \tau \pm r}
and so that $\sigma^-\vert_{\rm horizon} =0$,
for some time coordinate $\tau$. These coordinates pass continuously
through
the horizon, as shown in fig.~3.

As in two dimensions, the state is defined so that the in modes are in
their vacuum, and we wish to compare the resulting state to the out
vacuum.  The in coordinate with respect
to
which the vacuum is defined is $\xi^+$.  This is related to
$\sigma^+$,
then to $\sigma^-$, which is then related to $\xi^-$.
Let's
follow this chain backwards.  First, as $\xi^-\rightarrow\infty$, that is
near the horizon, $\sigma^-(\xi^-)$ is given by
\eqn\smxm{\frac{d\sigma^-}{d\xi^-} = \half
\left(\frac{\partial\sigma^-}{\partial t} -
\frac{\partial\sigma^-}{\partial r^*}\right) = \half
\left(\frac{\partial\tau}{\partial t} - \frac{\partial\tau}{\partial
r^*} + \frac{\partial r}{\partial r^*}\right)\ .}
This is simplified using
\eqn\fourfourteen{0=\frac{d\sigma^+}{d\xi^-} \propto
\frac{\partial\sigma^+}{\partial t} -
\frac{\partial\sigma^+}{\partial
r^*} = \frac{\partial\tau}{\partial t} - \frac{\partial\tau}{\partial
r^*}-\frac{\partial r}{\partial r^*}\ }
and we find
\eqn\fourfifteen{\frac{d\sigma^-}{d\xi^-} = \frac{\partial
r}{\partial
r^*}\ .}
Expanding about $r=r_{\rm horizon}$ then gives
\eqn\foursixteen{\frac{d\sigma^-}{d\xi^-} \simeq \frac{\partial
r}{\partial
r^*} \bigl(r=r_{\rm horizon}- \sigma^-/2\bigr) \simeq - \half
\ \frac{\partial}{\partial r}\left(\frac{\partial r}{\partial
r^*}\right)_{\Big|_{\rm horizon}}\sigma^-\ .}
The quantity
\eqn\fourseventeen{\kappa = \half\ \frac{\partial}{\partial r}
\left(\frac{\partial r}{\partial r^*}\right)_{\Big|_{\rm horizon}}}
is the usual surface gravity, and \foursixteen\ integrates to
\eqn\smrel{\sigma^- = -B\ e^{-\kappa\xi^-}}
for some constant $B$. This formula exhibits the expected
exponentially
increasing redshift near the horizon.  The coordinate $\sigma^-$ is
matched to $\sigma^+$ at $r=0$ and then $\sigma^+$ to $\xi^+$ at the
surface of the body.
For large $\xi^-$ we only
need
to relate $\sigma^-$ to $\sigma^+$ and
$\sigma^+$ to $\xi^+$ in the vicinity of the ray ${\cal
R}$ shown in fig.~3:
\eqn\fournineteen{\sigma^- \simeq{d\sigma^-\over d\sigma^+}
\frac{d\sigma^+}{d\xi^+}_{\Big|_{\cal
R}} \xi^+ +
{\rm const}\ .}
The factor $\frac{d\sigma^-}{d\xi^+}\Big|_{\cal R}$ gives a constant
blueshift,
and simply modifies $B$:
\eqn\sprel{\xi^+ \simeq B^\prime e^{-\kappa\xi^-}\ .}

This formula is nearly identical to \coordr.
The remaining derivation of the Hawking radiation follows
as in two-dimensions, with the
replacement $\omega\to \omega/\kappa$. Therefore the temperature is
\eqn\fourtwentyone{T=\frac{\kappa}{2\pi}\ ;}
for a Schwarzschild black hole this gives the familiar
\eqn\fourtwentytwo{T=\frac{1}{8\pi M}\ .}
The outgoing state is again thermal, except for the grey body
factors which are negligible for all but low-frequency modes.
Up to a numerical constant, the entropy of the Hawking radiation
is the same as that of the black hole, which is easily computed:
\eqn\fourtwentythree{dS = \frac{dE}{T} = 8\pi MdM\ ,}
so
\eqn\fourtwentyfour{S=4\pi M^2 = \frac{A}{4}\ ,}
where $A$ is the black hole area.  This is the famous
Bekenstein-Hawking
\refs{\Beke,\HawkEvap} entropy.

We can also easily estimate the lifetime of the black hole.  The
energy
density per out mode is $\omega/2\pi$; this is multiplied by the
thermal
factor, the velocity ($c=1$),
and the transmission coefficient $\Gamma_{\omega,\ell}$
through the barrier, then
summed
over modes:
\eqn\dmdt{\frac{dM}{dt} = \sum\limits_\ell (2\ell + 1) \int {d\omega\over
2\pi} \omega \frac{\Gamma_{\omega, \ell}}{e^{8\pi
M\omega}-1}\ .}
The transmission factor can be roughly
approximated by no transmission below the
barrier and unit transmission above,
\eqn\fourtwentysix{\Gamma_{\omega,\ell} \sim \Theta(a\omega M-\ell)}
for some constant $a$.  $M$ scales out of the integral \dmdt, and we
find
\eqn\fourtwentyseven{\frac{dM}{dt} \propto \frac{1}{M^2}\ .}
This gives a lifetime
\eqn\fourtwentyeight{\tau\sim \left(\frac{M}{\mpl}\right)^3 \tpl\ ,}
which is comparable to the age of the universe for $M\sim 10^{-18}
M_{\rm
sun}$.

Finally, note that as in two dimensions there are correlations
between
modes on either side of the horizon, and these mean missing
information
from the perspective of the outside observer.  Also, note that
although
the calculation refers to ultrahigh frequencies, in essence all that
is
being used is that the infalling state near the horizon is
approximately
the vacuum at high frequencies.

\newsec{Semiclassical treatment of the backreaction}

So far black hole shrinkage from Hawking emission has been neglected.
This section will treat
this backreaction in two dimensions and semiclassically;
for
attempts at construction of a full quantum theory of 2d black hole
evaporation see \eg\ \refs{\ChVe\SVV\StTh-\HVer}.

The effect of the Hawking radiation on the geometry is determined by
its
stress tensor. Recall that the asymptotic stress tensor was computed
in
sec.~3 by relating the normal ordering prescriptions in the two
different coordinate systems.  The metric in the out region asymptoted to
\eqn\fiveone{ds^2 = - d\xi^+ d\xi^- = - e^{2\rho} d\sigma^+
d\sigma^-\ .}
where
\eqn\fivetwo{e^{-2\rho} = \frac{d\sigma^+}{d\xi^+}
\ \frac{d\sigma^-}{d\xi^-}\ .}
With a little effort, \Tmm\ can be rewritten in terms of $\rho$.  Working
in $\sigma$ coordinates (but keeping track of which coordinates are used
for normal ordering) gives
\eqn\fivethree{:T_{--}:_\xi =
:T_{--}:_\sigma +  \frac{1}{12} \left[\partial ^2_-\rho - (\partial_-
\rho)^2\right] \equiv :T_{--}:_\sigma + t_{--}}
where we define $t_{ab}$ to be the difference between the two
normal-ordered
stress tensors.  This formula is valid for an {\it arbitrary} Weyl
rescaling $\rho$, and allows us to determine the full stress tensor
given the stress tensor in $\sigma$ coordinates.  In particular,
suppose
that the initial state satisfies
\eqn\fivefour{: T_{+-} :_\sigma = 0}
in accord with conformal invariance of the scalar field. Then the
conservation law
\eqn\fivefive{\nabla_+ T_{--} + \nabla_- T_{+-} = 0\ ,}
which should hold for the stress tensor in any generally coordinate
invariant regulation scheme, implies
\eqn\tcons{0=\frac{1}{12} \partial_+ \left[\partial^2_-\rho -
(\partial_-\rho)^2\right] + \partial_- t_{+-} - \Gamma^-_{--} t_{+-}
\ ,}
where we have used $\Gamma_{+-}^+ = \Gamma_{+-}^- =0$.
The Christoffel symbol $\Gamma_{--}^-$ is easily computed, giving
\eqn\fiveseven{\Gamma^-_{--} = g^{+-} \Gamma_{--,+} = 2\partial_-\rho
\ .}

Eq.~\tcons\ then integrates to
\eqn\fiveeight{t_{+-} = -\,\frac{1}{12}\, \partial_+\partial_-\rho\
,}
so
\eqn\coufan{: T_{+-} :_\xi = -\,\frac{1}{12}\, \partial_+\partial_-
\rho \ .}
The right side is proportional to the curvature $R$, and we have
\eqn\confanom{\langle T\rangle = {1\over 24} R\ .}
This is the famous {\it conformal anomaly}:
the regulated trace of
the stress tensor varies with the metric used to define the regulator.

Eq.~\coufan\ can be integrated to find the quantum effective action
of
the scalar field, using
\eqn\fiveten{\eqalign{\langle T_{+-} \rangle &= -\frac{1}{i}
\ \frac{2\pi}{\sqrt{-g}}\ \frac{\delta}{\delta g^{+-}}
\ \ell n\left[\int {\cal D}f\, e^{-\frac{i}{4\pi} \int (\nabla
f)^2}\right]\cr
&= -\frac{\pi}{4}\ \frac{\delta}{\delta\rho}\ S_{\rm eff}\ .}}
We find
\eqn\seff{S_{\rm eff} = -\,\frac{1}{6\pi} \int d^2\sigma\ \rho
\ \partial_+\partial_-\rho = -\,\frac{1}{24\pi}\ \int d^2 x
\ \sqrt{-g}\ \rho\ \sq\ \rho\ ,}
or, using the relation $R= -2\ \sq\ \rho$ and the Green function
$\sq^{-1}$ for
$\sq$,
\eqn\spl{\eqalign{S_{\rm eff} &= -\,\frac{1}{96\pi}\int d^2 x
\sqrt{-g}
 \int d^2x
\ \sqrt{-g^\prime}\ R(x)\, \sq^{-1}\, (x, x^\prime)\ R(x^\prime)\cr
&\equiv S_{\rm PL}}\ .}
This expression is known as the Polyakov-Liouville action\refs{\Poly}.

The quantum mechanics of the evaporating black hole is described by
the
functional integral
\eqn\fctint{\int {\cal D}g {\cal D}\phi\ e^{\frac{i}{\hbar}\, S_{\rm
grav}}
\int {\cal D}f\ e^{\frac{i}{\hbar}S_f} (\cdots)\ .}
Here we reinstate $\hbar$, and the ellipses denote matter sources
that
create the black hole.  In particular, if these are taken to be
classical, \fctint\ becomes
\eqn\fctintii{\int {\cal D}g {\cal D}\phi\ e^{\frac{i}{\hbar}
\, S_{\rm grav} + \frac{i}{\hbar}\, S^f_{c\ell} + i\, S_{PL}}\ .}
The Hawking radiation and its backreaction is encoded in $S_{PL}$,
which
corrects the functional integral at one loop.  However, there are
other
one loop corrections arising from $\phi$ and $g$, and the Hawking
radiation gets lost among them.  To avoid this, consider instead $N$
matter fields $f_i$, so \fctintii\ becomes
\eqn\Nint{\int {\cal D}g {\cal D}\phi\ e^{\frac{i}{\hbar}\, S_{\rm
grav}
+ \frac{i}{\hbar}\, S^f_{c\ell} + \frac{i}{\hbar}\ (N\hbar) S_{\rm
PL}}
\ .}
This has a semiclassical limit exhibiting Hawking radiation for
large $N$: $N\hbar$ must be fixed as $\hbar\to 0$. The
semiclassical
equations of motion are
\eqn\seom{\eqalign{G_{++} = T^{c\ell}_{++} + \frac{N}{12}
\left[\partial^2_+ \rho
- (\partial_+\rho)^2\right]&\cr
 - e^{-2\phi}
\left(2\partial_+\partial_-\phi - 4 \partial_+\partial_- \phi -
\lambda^2 e^{2\rho}\right)& \equiv G_{+-} = -\frac{N}{12}
\ \partial_+\partial_-\rho}}
and likewise for $G_{--}$; the dilaton equation is unmodified.

The equations \seom\ are no longer soluble. They have been studied
numerically \refs{\Lowe\PiSt-\Tada}, but instead we'll take a different
approach. Quantization of \Nint\ requires addition of
counterterms,\foot{For a general discussion of physical constraints on such
counterterms see \refs{\QTDG,\Japan}.} and
specific choices of these exist \refs{\BiCa\deAl-\RST} that
magically
restore solubility! These choices of counterterms don't spoil the
essential features of the evaporation, and the resulting theories are
thus soluble models for evaporating black holes.

We will focus on the prescription of Russo, Susskind, and
Thorlacius\refs{\RST}, who add counterterms that ensure the current
$\partial_\mu(\rho-\phi)$ is conserved at the quantum level.
To see how to accomplish this, add
\cgact\ and \seff\ to get the action including Hawking radiation effects,
\eqn\fiveseventeen{S_{sc} = \frac{1}{2\pi} \int d^2
x\left\{\left[2\nabla (\rho-\phi) \cdot \nabla\ e^{-2\phi}
+ 4
\,e^{2(\rho-\phi)}\right] + \frac{N\hbar}{12}\, \nabla\rho
\cdot \nabla\rho\right\}\ .}
Conservation of $\partial_\mu(\rho-\phi)$ can clearly be reinstated
by
adding the counterterm
\eqn\fiveeighteen{-\,\frac{\hbar N}{48\pi} \int d^2x\ \sqrt{-g}
\ \phi R = -\,\frac{1}{2\pi} \int d^2x\ \frac{N\hbar}{12}\ \nabla\phi
\cdot \nabla\rho\ ,}
and the action becomes
\eqn\srst{S_{\rm RST} = \frac{1}{2\pi} \int d^2x
\Bigl\{2\nabla(\rho-\phi)\cdot
\nabla \left(e^{-2\phi} + \frac{\kappa}{2}\ \rho\right) + 4
\,e^{2(\rho-\phi)}\Bigr\}\ .}
where we define $\kappa = N\hbar/12$.

Eq.~\srst\ is the same as \cgact\ with the replacement
\eqn\fivenineteen{e^{-2\phi} \to e^{-2\phi} + \kappa\rho/2\ ,}
so the solutions are found directly from \rhophi\ and \fmet,
\eqn\fivetwenty{\eqalign{\rho-\phi & = \half (\sigma^+ - \sigma^-)\cr
e^{-2\phi} + \frac{\kappa}{2} \rho & = M + e^{\sigma^+}
(e^{-\sigma^-} -
\Delta)\ .}}
Note in particular that
\eqn\fsc{e^{-2\phi} + \frac{\kappa}{2} \phi = M + e^{\sigma^+}
(e^{\sigma^-} -
\Delta) - {\kappa\over 4}(\sigma^+ - \sigma^-)\ .}
The left hand side has a global minimum at
\eqn\fcr{\phi_{\rm cr} = -\half\ \ell n(\kappa/4)\ ,}
so the solution becomes singular where the right hand side falls
below
this.  Here additional quantum effects should become important.  A
Kruskal diagram for the formation and evaporation is sketched in
fig.~5.

\ifig{\Fig\Kdiag}{Shown is a Kruskal diagram for collapse and evaporation
of a 2d dilatonic black hole.  The matter turns the
singularity at $\phi_{\rm cr}$ spacelike, and forms an apparent horizon.
The black hole then evaporates until the singularity and apparent horizon
collide.  The effective horizon is just outside the line $\sigma^- = - \ln
\Delta$.}{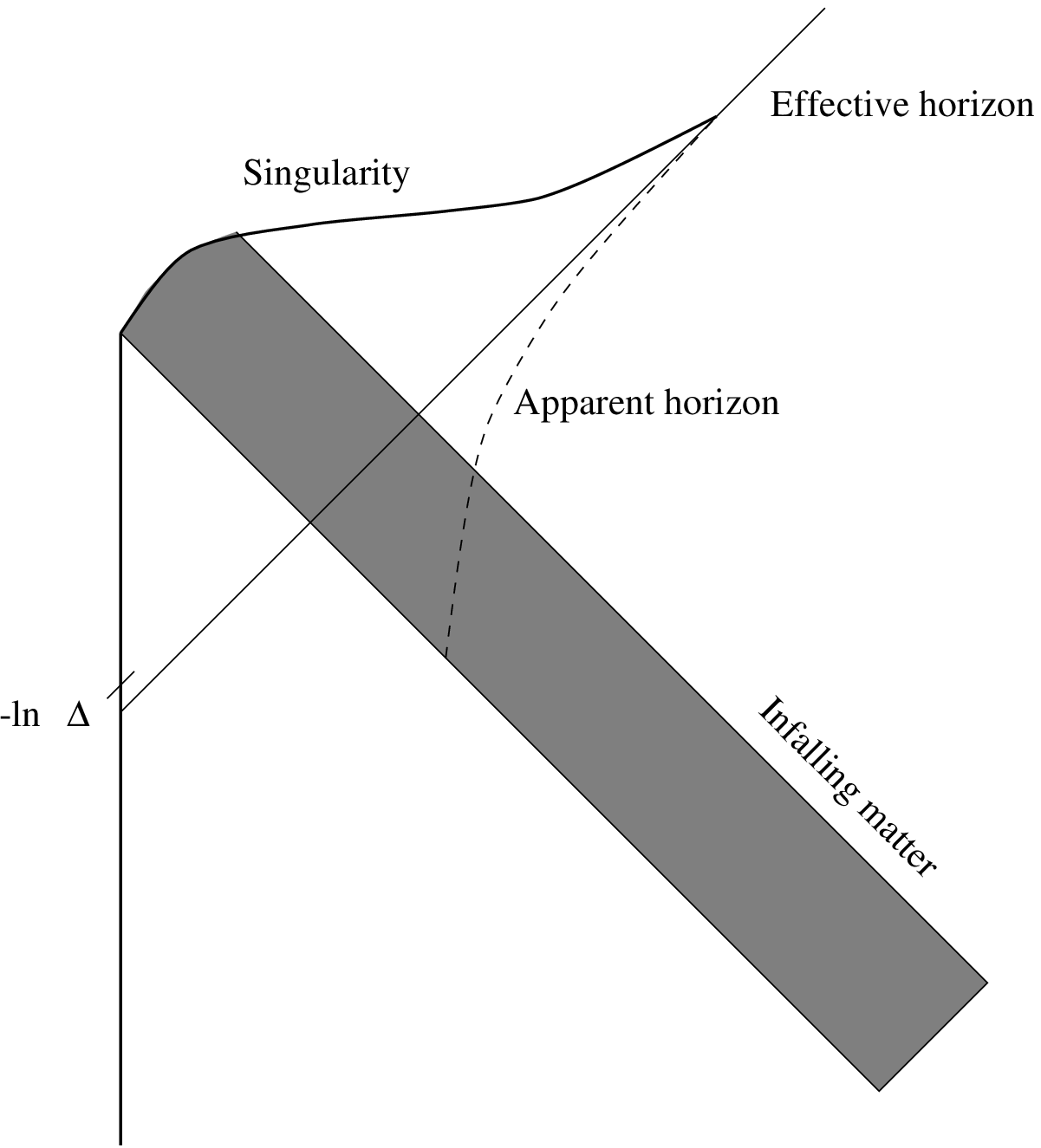}{3.50}

A sensible definition of the apparent horizon is the place where
lines
of constant $\phi$ become null, $\partial_+\phi = 0$ since if these
lines are spacelike it means one is inevitably dragged to stronger
coupling as in the classical black hole.  Differentiating \fsc,
the equation for the
horizon
is
\eqn\horloc{e^{-\sigma^-} = \Delta + {\kappa\over 4} e^{-\sigma^+}\ .}
The singularity becomes naked where it and the horizon meet,
\eqn\nsing{\sigma^-_{\rm NS} = -\ell n
\left(\frac{\Delta}{1-e^{-4M/\kappa}}\right)\ .}
With the singularity revealed, future evolution outside the black
hole
can no longer be determined without a complete quantum treatment of
the
theory: the semiclassical approximation has failed.  The last light
ray
to escape before this happens is called the effective horizon.  These
features are also shown in fig.~6.

\ifig{\Fig\penevap}{The Penrose diagram for the evaporating two-dimensional
black hole.  The lower part of the diagram has been truncated at $\phi_{\rm
cr}$, and the upper part  where the semiclassical
approximation fails due to appearance of large curvatures.}{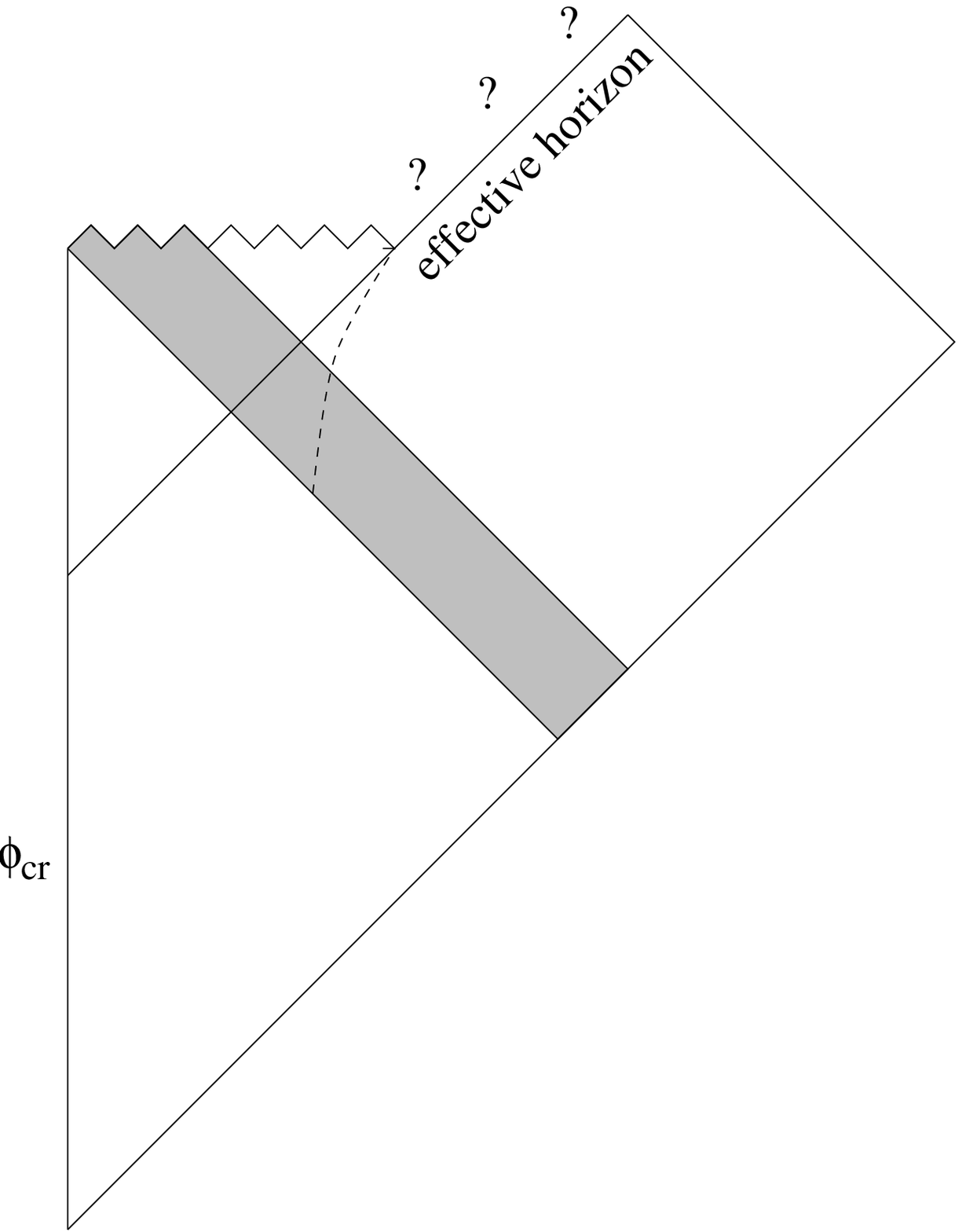}{3.50}

The Hawking flux is computed as before; as $\sigma^+ \to \infty$, the
metric asymptotes to the previous form \nine. The flux $\langle :
T_{--} :\rangle$ is therefore still given by \hawkst, and asymptotes
to
$\frac{1}{48}$ as $\xi^- \to \infty$.  As a check, we can compute the
mass lost up to the time \nsing\ where our approximations fail:
\eqn\fivetwentyfive{\int\limits_{-\infty}^{\xi^-_{\rm NS}}
d\xi^- \langle: T_{--} :\rangle =
M-\frac{\rm constant}{\kappa}\ .}
As anticipated, the semiclassical approximation breaks down when the
black hole reaches the analogue of the Planck scale, here $M_{bh}
\sim
1/\kappa$. Furthermore, as in the preceding  section it appears that
there are correlations between the Hawking radiation and the internal
state of the black hole, and that the Hawking radiation is
approximately
thermal.  An estimate of the missing information comes from the
thermodynamic relation
\eqn\fivetwentysix{dS = \frac{dE}{T}\ ,}
which gives
\eqn\fivetwentyseven{S=2\pi M\ .}

It was mentioned that attempts have been made to go beyond the
semiclassical approximation and define a complete quantum theory by
choosing a boundary condition\refs{\ChVe\SVV\StTh-\HVer},
for example reflecting,
at $\phi_{\rm cr}$. These attempts have met with various
obstacles, for example of instability to unending evaporation.
It is perhaps not surprising that a simple reflecting
boundary condition has
had difficulty in summarizing
the non-trivial dynamics of strong coupling.  Such a
boundary condition presumes that there are no degrees of freedom at strong
coupling, yet if one thinks of the connection of these solutions to
four-dimensional extremal black holes,
it seems quite possible that there are either a large number
of degrees of freedom or very complicated boundary interactions.

\newsec{The Information Problem}

\subsec{Introduction}

The preceding sections have outlined a semiclassical argument that
the
Hawking radiation is missing information that was present in the
initial
state.  We'd like to know what happens to this information. One
possibility was proposed by Hawking: the black hole disappears at the end
of evaporation and the information is simply lost.
Although on the face of it this is the most conservative solution, it
is
really quite radical; information loss is equivalent to a breakdown
of
quantum mechanics.

This prompts us to investigate other alternatives.  A second
possibility
is that the information is returned in the Hawking radiation.  This
could result from a mistake in our semiclassical argument involving a
fundamental
breakdown of locality and causality, as advocated in
\refs{\tHoo,\SVV,\HVer,\Suss}.
Another possibility is that the information is radiated after the
black
hole reaches $M\sim m_{pl}$ and the semiclassical approximation
fails.
Here ordinary causality no longer applies to the interior of the
black
hole, and it's quite plausible that the information escapes.

\ifig{\Fig\penrem}{A Penrose diagram appropriate to a long-lived
remnant scenario.
The singularity is replaced by a planckian region near
$r=0$.}{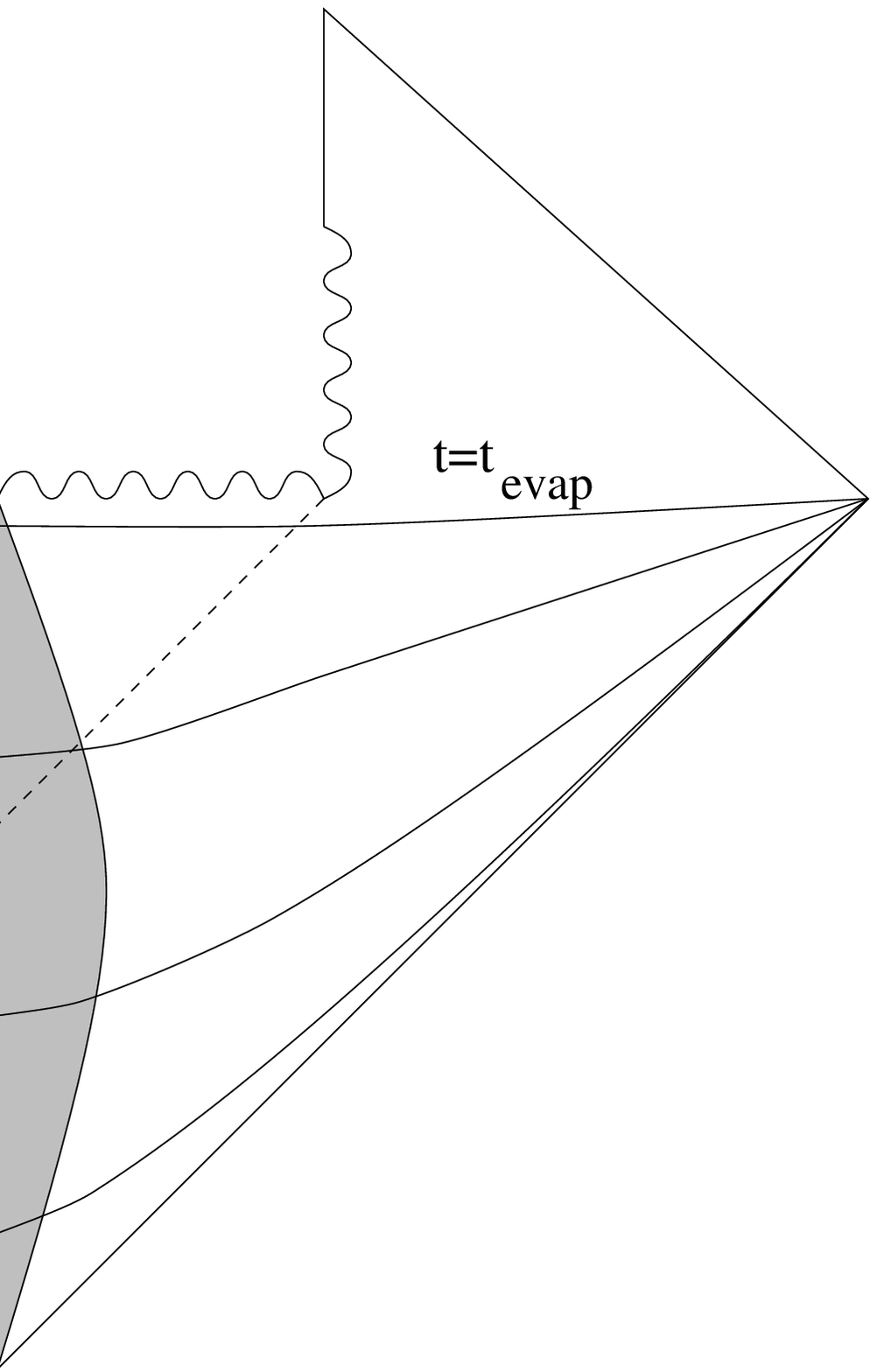}{3.00}

However, the amount of information to be radiated is, in four
dimensions,
given by $S\sim M^2$, and the amount of available energy is just the
remaining black hole mass, $M\sim\mpl$. The only way that the
outgoing
radiation can contain such a large amount of information with such a
small energy is if it is made up of a huge number of very soft
particles, for example photons. Each photon can carry approximately
one
bit of information, and so $M^2$ photons are required.  Their
individual
energies are therefore $E_\gamma \sim 1/M^2$, and the decay time to
emit one such photon is bounded by the uncertainty principle,
\eqn\sixone{\tau_\gamma \roughly>\frac{1}{E_\gamma} \sim M^2\ .}
It then takes a time
\eqn\sixtwo{\tau_{\rm rem} \sim \left(\frac{M}{\mpl}\right)^4 \tpl}
to radiate all of the information in $M^2$ photons\refs{\CaWi,\Pres}.
For example, this approximates the
lifetime of the universe for a black hole whose initial mass is that
of
a typical building.  This implies the third alternative: that the
black hole
leaves a long-lived, or perhaps stable, remnant.  An example is
exhibited in the Penrose diagram of fig.~7: after the Hawking
radiation
ceases, a remnant is left at the origin.  A qualitative picture of
such
an object is shown in fig.~8, which shows the geometry of the
time slice
$t_{\rm evap}$ in fig~7. The remnant is a long planckian
fiber with radius
$\sim\rpl$, and the infalling matter concentrated at its tip.
Describing such a configuration certainly requires planckian physics,
and it is quite plausible that this physics allows the remnant to
slowly
decay.

\ifig{\Fig\remsli}{A late time slice through fig.~7. shows a long
Planck sized fiber attached to an asymptotically flat
geometry.}{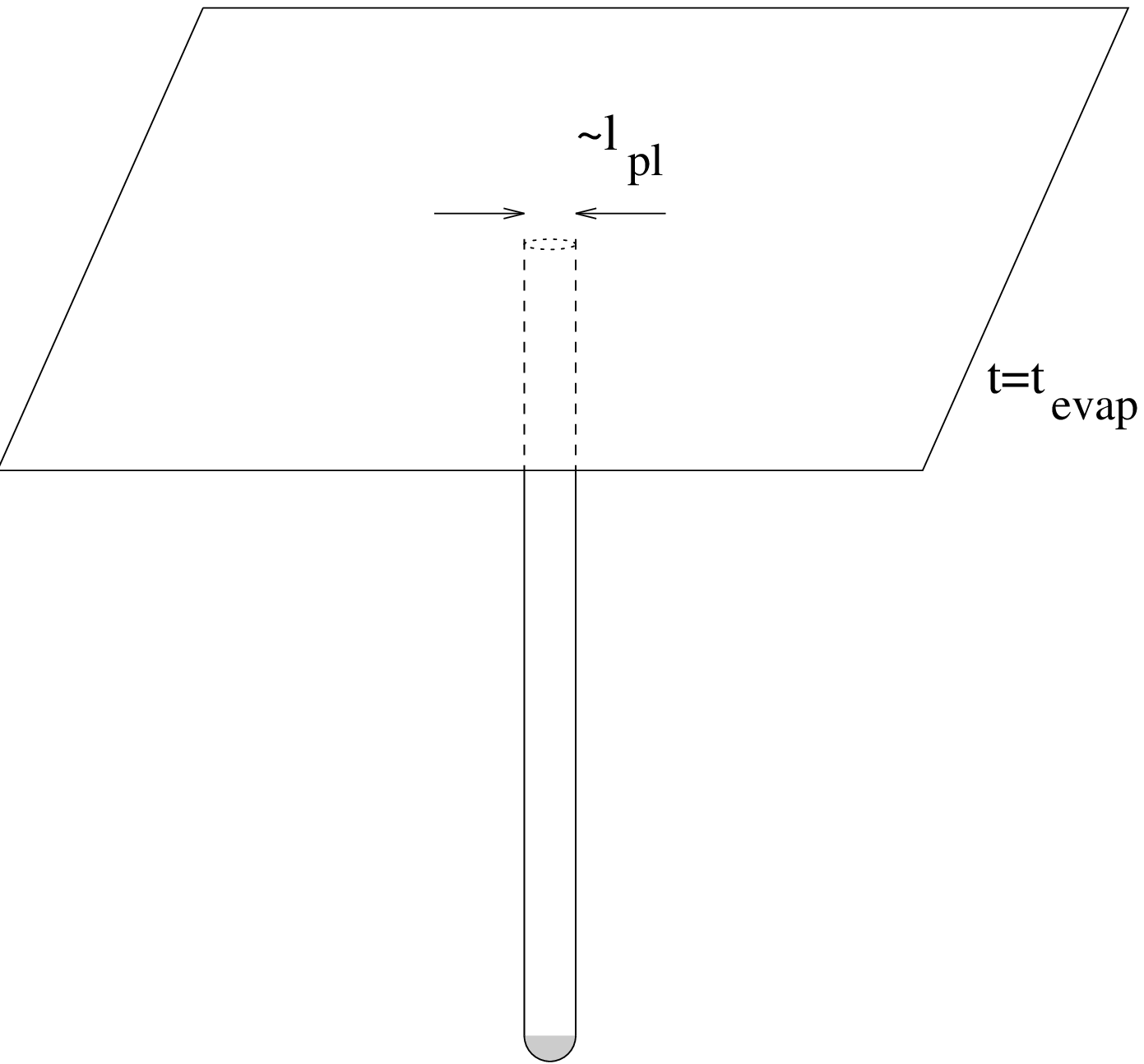}{2.00}

Each of these three scenarios has staunch advocates.  Each also
appears
to violate some crucial principle in low-energy physics.  The
resulting
conflict is the black hole information problem.  Let's investigate
this
in more detail.

\subsec{Information Loss}

Information loss violates quantum mechanics, but worse, it appears to
violate energy conservation, and this is a disaster.  The heuristic
explanation for this is that information transmission requires
energy,
so information loss implies energy non-conservation.  Suppose for
example that we attempt to summarize the formation and evaporation
process by giving a map from the initial density matrix to the final
outgoing density matrix,
\eqn\sixthree{\rho_f =\not\kern-0.2em S \rho_i\ ,}
where $\not\kern-0.2em S$ is a linear operator. Such operators, if
not
quantum mechanical,
\eqn\sixfour{\rho_f \not= S\,\rho_i S^+\ ,}
typically either violate locality or energy conservation
\refs{\BPS\Sred}.

A useful way of explaining
the connection between information and energy
is to model the black hole
formation
and evaporation process as the interaction of two Hilbert
spaces\refs{\CILAR},
 ${\cal H}_o, {\cal H}_h$.  The first includes the states
of the outside world, and the second the internal states of the black
hole which are thought of as ``lost''. Suppose these interact via a
conserved hamiltonian
\eqn\thamil{H= H_o + H_i + H_h}
where $H_o$ acts only on ${\cal H}_o$, $H_h$ only on ${\cal H}_h$,
and
the interaction hamiltonian $H_i$ acts on both. $H_i$ summarizes the
physics that transfers information from the outside Hilbert space to
the
hidden one.

Now, the black hole formation and evaporation process involves loss
of
information during a time $\tau \sim M^3$. On larger time scales the
process can be repeated at the same location;
that is, another independent black hole
formation and evaporation can take place.  In general, we say that the
information
loss is {\it repeatable} if the information loss in $n$ such
experiments is $n$ times the information loss in a single experiment
\eqn\sixsix{\Delta S_n = n\Delta S_0\ .}
One can plausibly argue \refs{\CILAR} if not prove that repeatable
information loss on a time scale $\Delta t$ indicates energy losses
\eqn\sixseven{\Delta E \propto 1/\Delta t}
from the observable Hilbert space (our world) to the hidden one (the
interior of the black hole). Suppose, for example, that there is no
energy loss,
$H_h=0$.  A simple example for the hidden Hilbert space is the states
of
a particle on a line, ${\cal H}_h = \{|x\rangle\}$, and an example
hamiltonian is
\eqn\sixeight{H_i = {\cal O}\hat x}
for some operator ${\cal O}$ acting on the observable space.  The
wavefunction $|\Psi\rangle$ of the coupled system obeys the
Schr\"odinger
equation with hamiltonian \thamil, and the outside density matrix is
defined as
\eqn\sixnine{\rho_o = tr_h  |\Psi\rangle\langle\Psi|\ .}
One can readily show \refs{\herring} that in $n$ scattering
experiments,
the information loss per experiment declines:
\eqn\sixten{\lim\limits_{n\to\infty} \frac{\Delta S_n}{n} = 0\ .}
This is not repeatable information loss.

The explanation for this is familiar from wormhole physics
\refs{\herring,\LInc}. If we started in an $\hat x$ eigenstate,
\eqn\sixeleven{|\Psi\rangle = |\Psi_0\rangle|x\rangle\ ,}
then the sole observable effect of the interaction
is a modification of the external
hamiltonian:
\eqn\sixtwelve{H= H_o + {\cal O} x\ .}
The eigenvalue $x$ simply plays the role of a coupling constant.  If
we
instead started in a superposition of $\hat x$ eigenstates, then we
simply have a probability distribution for coupling constants.
For example, suppose we begin in the superposition
\eqn\superpos{\alpha|x_1\rangle +\beta |x_2\rangle\ . }
The hamiltonian does not change this internal state; there is a
superselection rule.  If we compute the expectation value of some local
operators $\calO_1,\cdots,\calO_n$ in this state, it gives
\eqn\exptval{|\alpha|^2 \langle x_1| \calO_1\cdots\calO_n |x_1\rangle +
|\beta|^2 \langle x_2| \calO_1\cdots\calO_n |x_2\rangle\ .}
Performing repeated experiments simply
correlates the outside state with
the
value of the effective coupling constant $x$, that is, measures the coupling
constant. Once these are determined there is no further loss of
information.

These arguments generalize to more interesting Hilbert spaces ${\cal
H}_h$, and more general interaction hamiltonians,
\eqn\sixthirteen{H_i = \sum\limits_\alpha {\cal O}_\alpha {\cal
I}_\alpha}
where ${\cal O}_\alpha$ and ${\cal I}_\alpha$ act on ${\cal H}_o$ and
${\cal H}_h$ respectively.  Finally, note that if $H_h \not= 0$ but
the
operators ${\cal I}_\alpha$ only act between energy levels with
energy
separation $\ltwid \Delta E$, then the information loss will not be
repeatable on time scales $\Delta t \ltwid \frac{1}{\Delta E}$. This
indicates a basic connection between information loss and energy
non-conservation.

\ifig{\Fig\evbh}{Shown is the contribution of a virtual black hole to two
body scattering; information and energy loss are expected to contaminate
arbitrary processes through such diagrams.}{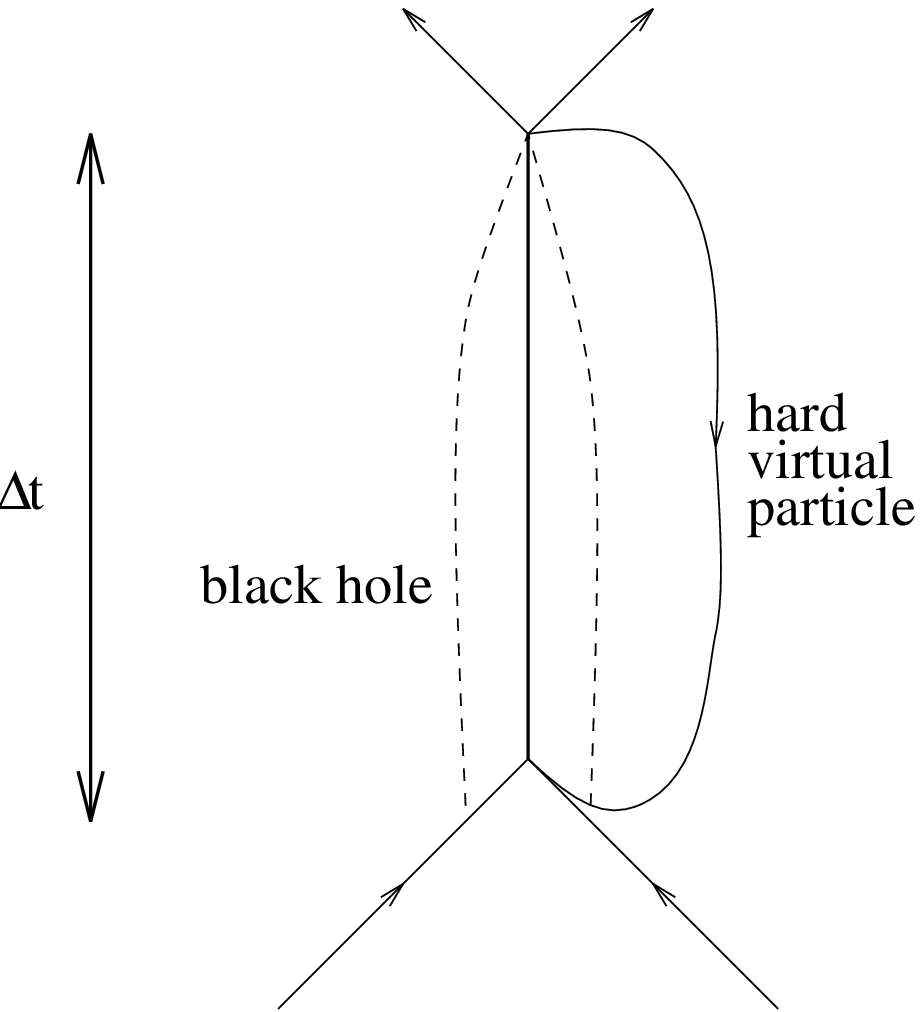}{2.00}

Worse still, by basic quantum principles, if information and energy
loss is allowed, it will take place in virtual processes. An example
is
shown in fig.~9. For such a virtual process on time scale $\sim
\Delta t$ one expects energy loss $\Delta E \sim 1/\Delta t$.
Furthermore, there is no obvious source for suppression from such
processes occurring at a rate of one per
Planck
volume per Planck time with $\Delta t\sim\mpl$. Allowing such energy
fluctuations is analogous to putting the world in contact with a heat
bath at temperature  $T\sim T_{\rm pl}$: this is in gross
contradiction
with experiment.

Indeed, \refs{\BPS}\ argued that a general class of local
$\Ssl$ matrices mimic such thermal fluctuations.  It can be shown
\refs{\CILAR}
that these correspond to a thermal distribution at infinite
temperature.

Attempts to turn Hawking's picture of information loss into a
consistent scenario are found in \refs{\PoSt} and very recently
in \StU, where a theory of
long-lived remnants emerges (see section 6.4).

\subsec{Information return}

Our semiclassical description, together with the assumption that the black
hole disappears, clearly led to information loss --- the
Hawking radiation at any given time was thermal with no correlations
with either the infalling matter or with earlier Hawking radiation.
Furthermore, the semiclassical approximation appears valid up until
the
time when the black hole reaches the Planck mass; before this time
curvatures are everywhere small. In the two-dimensional model of
section
2, the rate of mass loss of the black hole is proportional to $N$,
and
using the relation \fivetwentyseven\
between information and energy, one finds that
if
the information is to be radiated before the Planck scale it must be
at
a rate
\eqn\sixfourteen{\frac{dI}{dt} \propto N}
by the time the black hole has reached a fraction of its original
mass,\foot{For a related argument that information return must begin
by
this time see \refs{\Page}.} say $M/10$. No such information return
at
this order in the large-$N$ expansion is seen.

In fact, there is another argument that the Hawking radiation
doesn't contain information.  For it to do so, the outgoing state
would
have to be some modification of the Hawking state.  If we evolve such
a
state backwards in time until it reaches the vicinity of the horizon,
this state will differ from the local vacuum, and because of the
large
redshift the difference will be in very high energy modes.  An
infalling
observer would encounter these violent variations from the vacuum at
the
horizon.  This conflicts with our belief that there is no local way
for
a freely falling observer to detect a horizon: we could all be
falling
through the horizon of a very large black hole at this very moment,
and
may not know it until we approach the singularity.

\ifig{\Fig\locslic}{Locality in field theory implies that observations made
at $x$ and $y$ with spacelike separations must commute.}{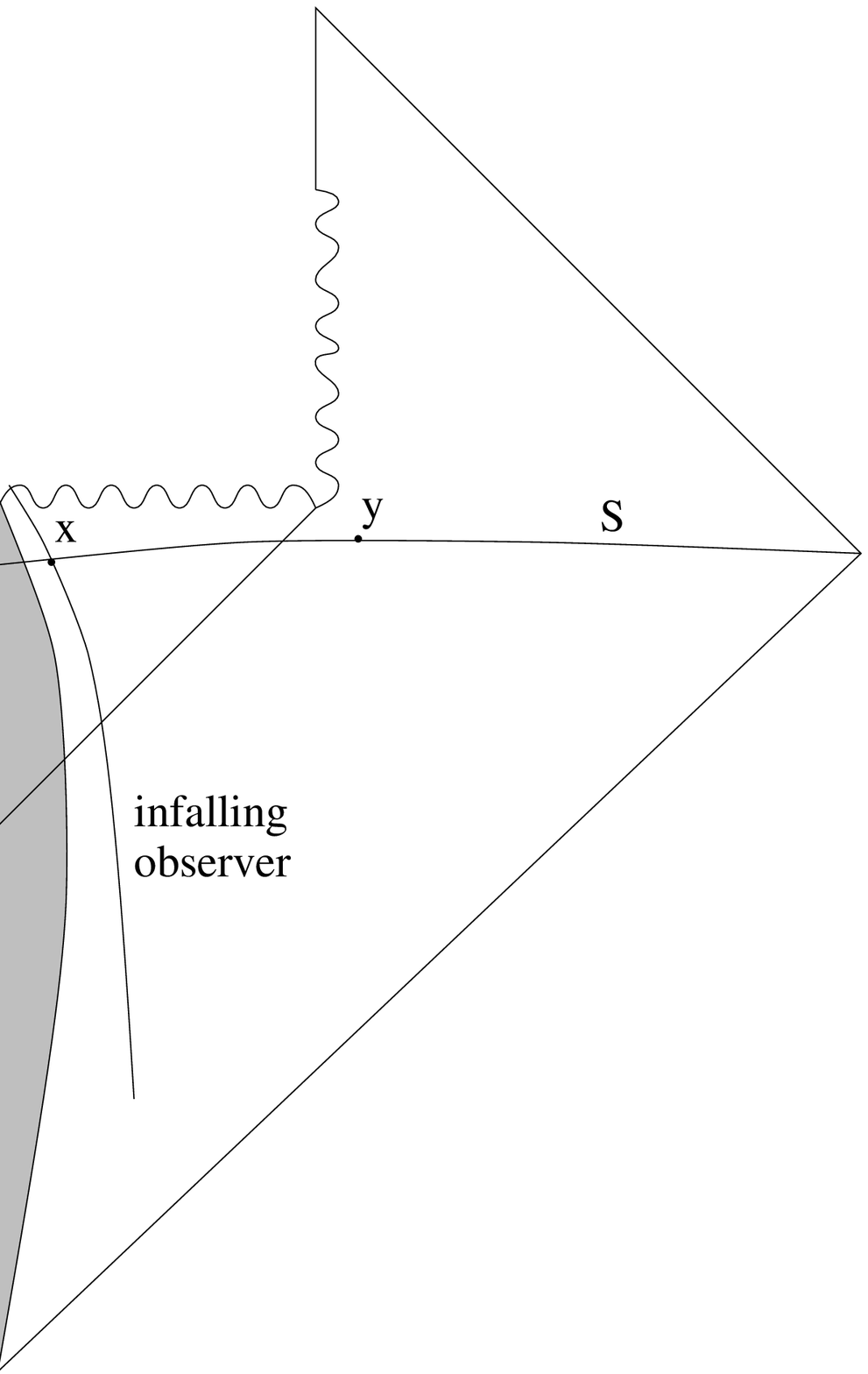}{3.50}

That information is not returned is a consequence of locality and
causality.  To see this, consider the spacelike slice $S$ of fig.~10.
Locality/causality in field theory is the statement that
spacelike-separated local observables commute, so
\eqn\locality{[{\cal O}\,(x), {\cal O}\, (y)] = 0\ .}
This means that up to exponentially small tails, the state on the
slice
$S$ can be decomposed
\eqn\sixsixteen{\psi ``=" \sum\limits_\alpha |\psi_\alpha\rangle_{\rm
in}
\ |\psi_\alpha\rangle_{\rm out}}
where $|\psi_\alpha\rangle_{\rm in}$, $|\psi_\alpha\rangle_{\rm out}$
are states with support inside or outside the black hole.  As we've
said, infalling observers don't encounter any particular difficulty
at
the horizon, and so their measurements reveal different internal
states
$|\psi_\alpha\rangle_{\rm in}$ depending on the details of the
collapsing matter, etc. Thus when we trace over internal states to
find
the density matrix relevant to the outside observer, it has missing
information, $\Delta I \sim M^2$.  The only obvious way one could
avoid correlations leading to this
impurity is if the internal state of the black hole is {\it unique},
\ie\
\eqn\bleach{\psi=|\psi_0\rangle_{\rm in} |\psi\rangle_{\rm out}\ ,}
and in particular
is independent of the infalling matter.  In this case an infalling
observer
would be ``bleached'' of all information when crossing the horizon.  Since
there should be no way to discern a horizon locally this appears impossible.

A possible weakness of this argument is the notorious problem of
defining observables in quantum gravity;\foot{For related discussion
in
two-dimensional models, see \refs{\HVer}.} Perhaps there are no truly
local observables. However, within regions where the semiclassical
approximation is valid one expects the observables of quantum gravity
to
reduce to ordinary field theoretic observables, up to small
corrections.

An extreme example of what information return in the Hawking
radiation
entails comes from the observation that the vicinity of the horizon
for
a large mass black hole is well approximated by flat space.  Indeed,
in
the limit $|r-2M| << 2M$, the Schwarzschild metric
\eqn\sixseventeen{ds^2 = -\left(1-\frac{2M}{r}\right) dt^2 +
\frac{dr^2}{\left(1-\frac{2M}{r}\right)} + r^2 d\Omega^2_2\ }
is well approximated by
\eqn\Mlarge{ds^2= -\ \frac{x^2}{16M^2} dt^2 + dx^2 + 4M^2 d\Omega^2_2
\ ,}
as can be shown using the substitution $x^2=8M(r-2M)$.
For large $M$, $4M^2 d\Omega^2_2 \simeq dy^2 + dz^2$, and the line
element \Mlarge\ is that of flat Minkowski space as seen by a family
of
accelerated observers.  This is known as Rindler space, and is
shown
in fig.~11. If the information from the infalling matter is in
the
Hawking radiation, it must be accessible to observers outside the
horizon, and this should also hold true as $M\to\infty$.  Then
information about states in the entire left half of Minkowski space is
accessible to observers in the right wedge of Rindler space.  If I
walk
past the horizon and keep going for a billion light-years to point
$x$,
the observer at $y$ still has full access to my internal state.  This
represents a gross violation of the locality/causality, \locality.

\ifig{\Fig\trsli}{The right wedge of Minkowski space corresponds to Rindler
space; it is the region observable by uniformly accelerated observers.
Taking the $M\rightarrow \infty$ limit of the black hole nonlocality
arguments shows that information at an arbitrary point x is fully
accessible in the vicinity of the Rindler horizon.}{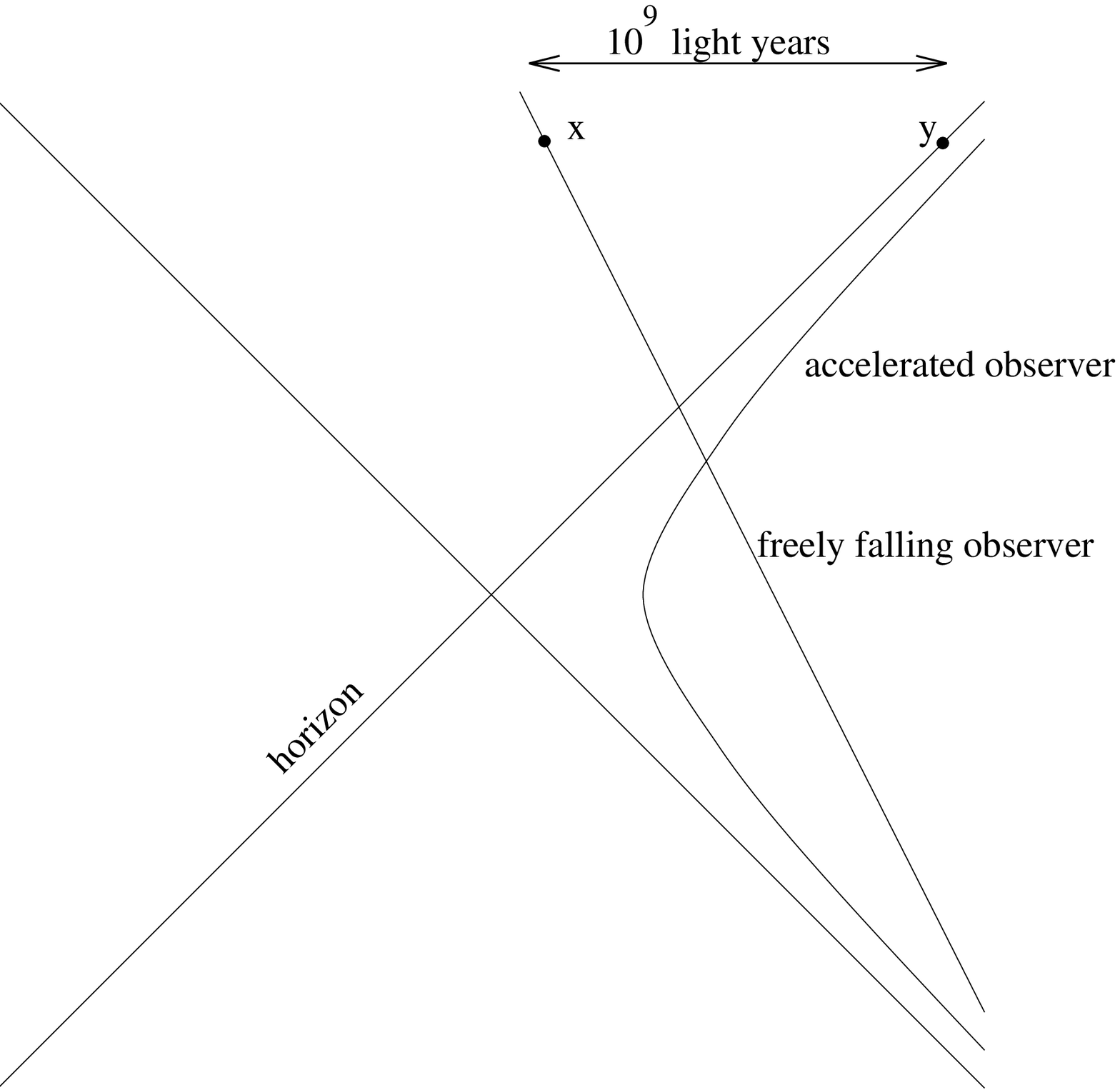}{3.25}

This hypothetical ``holographic'' property of spacetime, that
all
information in a three-dimensional region is encoded on the surface
of
that region,
requires drastically new physics and has been advocated by
't Hooft\refs{\tHoo} and more recently by Susskind, Thorlacius, and Uglum
\refs{\Suss}. 't
Hooft
efforts involve the hypothesis that the world is at a fundamental
level
similar to a cellular automaton\refs{\tHooaut}.
Susskind et. al.'s attempts instead rely on
string
theory.

Indeed, unlike point particles, strings are intrinsically non-local
objects.  Measurements made in the vicinity of points $x$ and $y$ of
fig.~11 in string theory are expected not to commute.  But these
violations
of locality are expected to generically fall off exponentially fast
on
the string scale, which is approximately the Planck length,
\eqn\sixnineteen{\left[{\cal O}\,(x), {\cal O}\, (y)\right] \sim
e^{-(x-y)^2/\ell^2_{st}}\ ,}
and as such are too small to be relevant to the information problem.
However, the Rindler observers are moving at huge velocities relative
to
those freely-falling, and Lowe, Susskind, and Uglum \refs{\LSU} argue
that these enormous relative boosts compensate for the exponential
fall
off and make
\eqn\modcom{\left[{\cal O}\,(x), {\cal O}\,(y) \right] \sim 1}
for suitable ultra high-energy observations at $y$.

So in string theory, it is possible that in some sense the
information
from the left half of the world is encoded in the right half.
In fact, all of the information in the Universe might be expected to be
encoded in the surface of a piece of chalk, a post-modern take on
Huxley\Hux!
However,
an open question is whether it is encoded in any realistically
accessible fashion --- \modcom\ would hold only for particular ultra
high-energy observations, and it is not clear that the information
available from such unrealistic observations is imprinted on the
Hawking
radiation. At present no detailed mechanism to transmit the information
to the
Hawking radiation has been found. In fact, a logical possibility
is
that the information never escapes from the horizon, and this is
 consistent with a remnant picture.

In conclusion, information return requires violations of
locality/causality. Since locality and causality are at best poorly
understood in string theory, this opens the  possibility of
information
return. However, an explicit
quantitative picture of how this happens is yet
to be produced.

\subsec{Remnants}

If black hole remnants, either stable or long lived,
resolve the information problem, then there
must
be an infinite number of remnant species to store the information
from
an arbitrarily large initial black hole.  For neutral black holes
Hawking's calculation fails at the Planck mass, so this means an
infinite species of Planck-mass particles.  These
are expected to be infinitely pair produced in generic physical
processes --- a clear disaster.

This statement is most easily illustrated if we imagine that remnants
carry electric charge; we return to the neutral case momentarily.
Stating that a remnant is charged is equivalent to assuming that
there
is a non-zero minimal-coupling to low-frequency photons, as shown in
fig.~12.  By crossing symmetry, this coupling implies that pairs will
be
produced by Schwinger production\refs{\Schw} in a background electric
field. The total production
rate is
\eqn\sixtwentyone{\Gamma_{\rm vac} \sim N\ e^{-\pi m^2/qE}}
where $m$ is the remnant mass, $E$ is the field strength, and
$N$ is the number of species.  Although the exponential may be
small it is overwhelmed in the case of infinite species.

\ifig{\Fig\Cross}{Using crossing symmetry, coupling of a particle to the
electromagnetic field implies Schwinger pair production.}{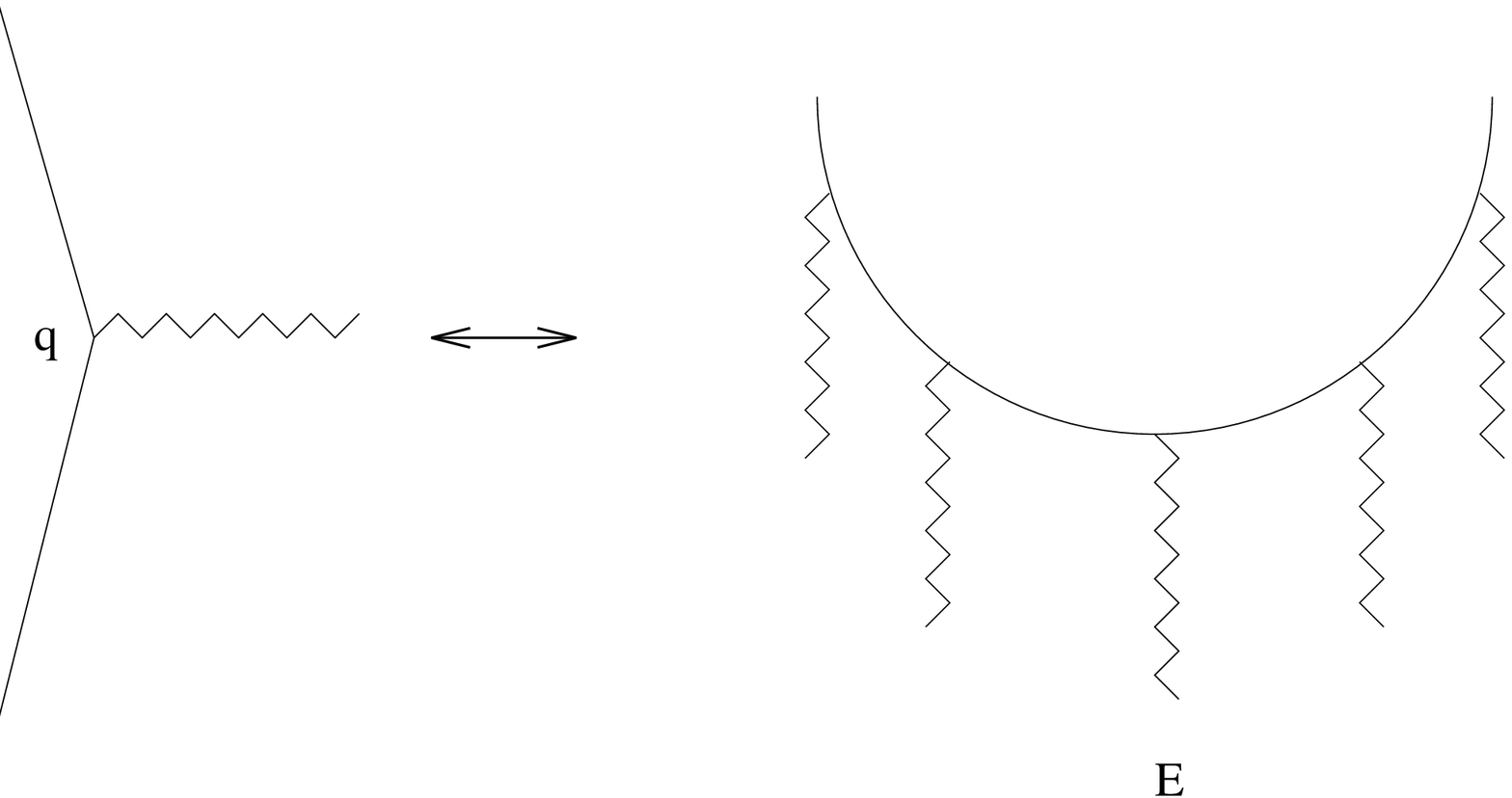}{1.8}

Neutral remnants have the same problem.  For example, the
gravitational
analog of Schwinger production is Hawking radiation, for which we'd
find
a rate
\eqn\sixtwentytwo{\frac{dM}{dt} \propto N\cdot \frac{1}{M^2}}
for a black hole to decay. $N=\infty$ gives an infinite rate.
Likewise
pairs could be produced in other everyday physical processes that have
sufficient available energy.  Although such production is highly
suppressed by small form factors, this suppression is overcompensated
by
the infinite states.

\subsec{Summary}

\begintable
proposal|principles violated\cr
information loss|unitarity, energy conservation\crnorule
information radiated|locality/causality\crnorule
remnants|crossing symmetry\endtable
\centerline{Table 1}

A summary of the proposed fates of information is shown in Table 1,
along with the corresponding objections. Each
of these objections can be phrased solely in terms of low-energy
effective physics, and
is apparently independent of any hypothesized
Planck-scale physics.  For this reason this conflict
has
been called the information {\it paradox}. The matter would be
reduced
to merely a problem if a convincing way were found whereby
Planck-scale
physics might evade one of these objections.

\newsec{Black hole pair production}

An aspect of the quantum mechanics of black holes that is of interest
in
its own right is the pair production of black holes.  This phenomenon
is
also of direct relevance to the information problem.  Indeed, let us
assume the validity of unitarity (no information destruction) and
locality (no information return in Hawking radiation). We've seen
that
these imply evaporating black holes leave an infinite variety of
neutral
remnants.  The same reasoning in the charged sector implies that
there
are an infinite number of internal states of an extremal, $M=Q$, \RN\
black hole. Indeed, assume we begin with an extremal hole; we may
then
feed it a huge amount of information, say, by dropping in the planet
Earth.  The black hole then Hawking radiates back to $M=Q$ and leaves
an
extremal black hole with the extra information encoded in its
internal
states.  This process may be continued indefinitely, implying an
infinite number of internal states of \RN\ black hole.

Infinite states suggest unphysical infinite Schwinger
pair production.  There
are
two possibilities.  The first is that a careful calculation of the
production rate gives a finite answer.
In this case \RN\ black holes would provide
an example of how to avoid the infinite states/infinite production
connection that could likely be translated into a theory of neutral
remnants with finite production.\foot{Dilaton black holes have also been
investigated in this context\refs{\BaOl,\BOS}.}
Alternatively, the assumption of
infinite
states may indeed imply infinite production.  This then
would
appear to rule out either unitarity or locality, and remove the
raison
d'\^etre for neutral remnants.

An instanton exists describing such pair production, but let us first
discuss this process following Schwinger's original arguments.  Since
\RN\ black holes --- or neutral remnants --- are localized objects
and
must have a Lorentz invariant, local and causal, and quantum
mechanical
description, they should be described by an effective quantum
field\refs{\CBHR,\CILAR}.  This field should have a species label,
\eqn\sevenone{\phi_a(x)\ ,}
and let these have masses $m_a$.
We will  take
these to be electrically charged with charge $q$, although the case
of magnetic monopole production in a magnetic field is equivalent by
electromagnetic duality.  This means that there should be a minimal
coupling
interaction with the electromagnetic field, plus higher dimension
operators:
\eqn\efacn{S_{\rm eff} = \int d^4 x \sum\limits_a \left(-|D_\mu
\phi_a|^2 - m^2_a|\phi_a|^2\right) + \cdots}
with $D_\mu = \partial_\mu + iq\, A_\mu$.

The decay rate into species $a$ of the vacuum consisting of a
background
electric field with vector potential $A^\mu_0$ is given
by
the vacuum-to-vacuum amplitude,\foot{See, \eg, \ItZu.}
\eqn\seventhree{e^{-\frac{\Gamma_a}{2}\,VT-iE_0T} = \langle
0|0\rangle_{A_0} = \frac{\int {\cal D}\tilde A_\mu {\cal D}\phi_a
\ e^{iS_{\rm eff} \left[A_0 + \tilde A, \phi_a\right]}}{\int {\cal D}
\tilde A_\mu {\cal D}\phi_a\ e^{iS_{\rm eff} \left[\tilde A,
\phi_a\right]}}\ ,}
where $V$ is the volume, and where
we have separated off electromagnetic fluctuations $\tilde A$.
To
leading order in an expansion in the charge, this gives
\eqn\sevenfour{\eqalign{\Gamma_a \cdot VT &= -2 Re\ \ell n\
Det^{-\half}
\left(\frac{-D^2_0 + m^2_a}{-\partial^2 + m^2_a}\right)\cr
&= Re\ Tr\ \ell n \left(\frac{-D^2_0 + m^2_a}{-\partial^2 +
m^2_a}\right)\ .}}

\ifig{\Fig\tschw}{Above the slice $S$ (dotted) is shown the lorentzian
trajectories of a pair of oppositely charged particles in an electric
field.  Below $S$ is the euclidean continuation of
this solution.  Matching the euclidean and lorentzian solutions smoothly
along $S$ gives a picture of Schwinger production followed
by subsequent evolution of the pair of created particles.}{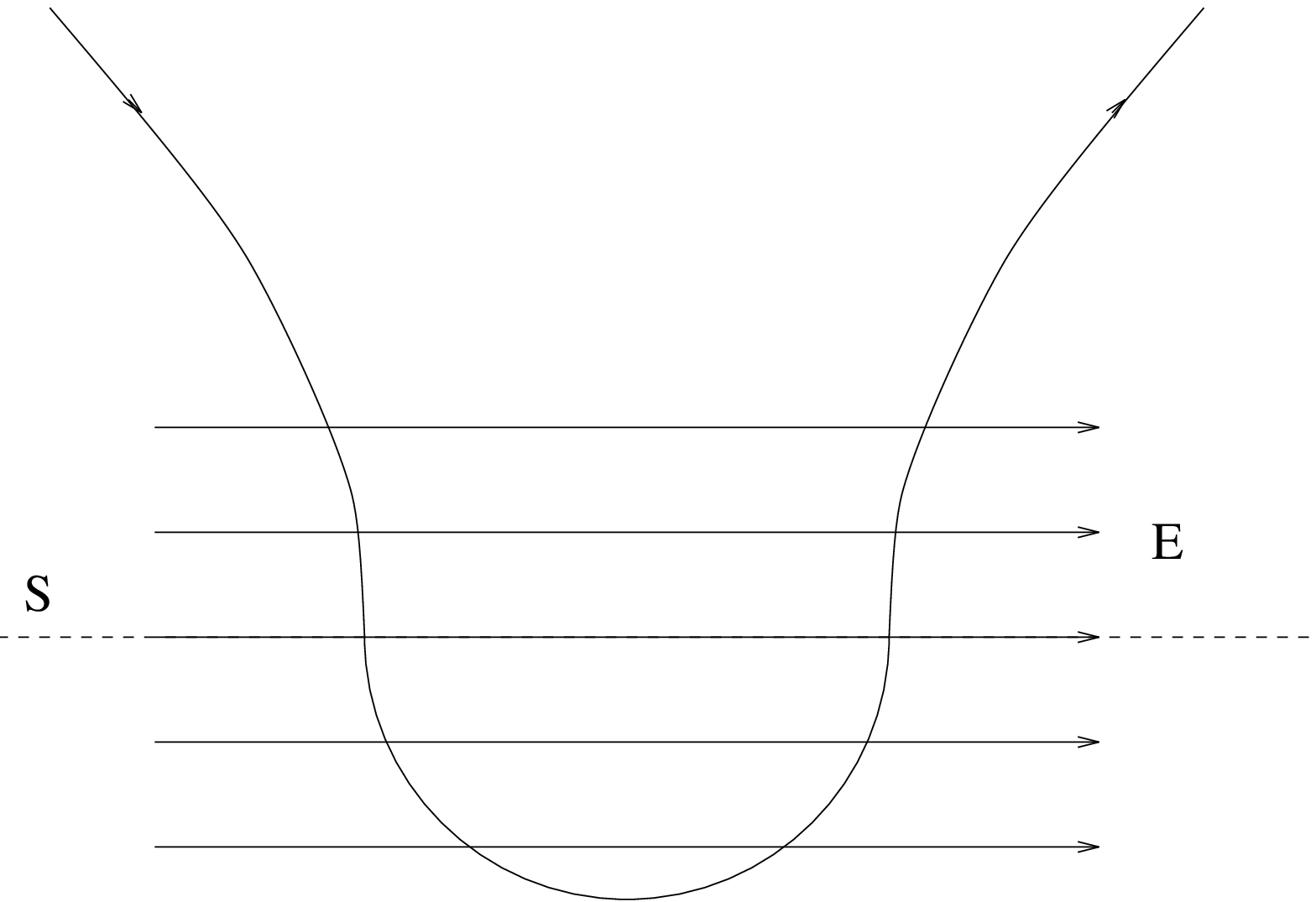}{1.75}

The trace can be rewritten in terms of a single-particle amplitude:
\eqn\sevenfive{\Gamma_a\cdot VT = {\rm Re}
\int d^4 x \int\limits^\infty_0
\frac{dt}{t} \left(\langle x| e^{iH(A^0)t} | x\rangle - \langle x |
e^{iH(0)t} |x\rangle\right)}
with
\eqn\sevensix{\langle x^\prime | e^{iH(A)t} | x\rangle =
\int\limits^{x^\prime}_x {\cal D} x\ e^{i\int^t_0
d\tau\left(\frac{\dot
x^2}{2} + iq A_\mu \dot x^\mu + \frac{m^2_a}{2}\right)}\ .}
This functional integral has a dominant euclidean saddlepoint
corresponding to circular motion of the particle in the field.
This instanton is the euclidean continuation of a solution
where
a pair of charged particles started at rest run hyperbolically to
opposite ends of the background field, as shown in fig.~13.  Then if
the
euclidean trajectory is cut on the dotted line, where the velocities
vanish, it smoothly matches to the lorentzian solution: pairs are
created and run to infinity.

The action of the instanton is easily found to be
\eqn\sevenseven{S_E = \frac{\pi m^2_a}{qE}\ ,}
giving an approximate decay rate into species $a$,
\eqn\seveneight{\Gamma_a \sim e^{-\pi m^2_a/qE  }+\cdots\ .}
Corrections arise both from the higher order terms in \efacn\ and
from
subleading quantum corrections. These are typically subleading in an
expansion in $qE$. The total decay rate is
\eqn\sevennine{\Gamma \sim \sum\limits_a e^{-\pi m^2_a/qE}\ .}
An infinite spectrum of remnants with nearly degenerate masses
clearly
gives an infinite decay rate.  Note in particular that if the spectrum is
of the form $m_a=m_0 +\Delta m_a$, where $\Delta m_a \ll m_0$ are small
mass splittings, then  the decay rate is proportional to the partition
function for internal states:
\eqn\proptopart{\Gamma\sim e^{-\pi m^2_0/qE}\, {\rm Tr}_a
e^{-\beta \Delta m}\ ,}
with $\beta=2\pi m_0/qE$.

Turning now to black holes, the remarkable fact is that solutions
analogous to those in fig.~13 are explicitly known
\refs{\Erns\Gibb\GaSt\DGKT-\DGGH} for charged black holes!
Although
the solutions are known for arbitrary strength coupling to a
dilatonic
field (as in string theory), here we'll consider only the simpler
case
without a dilaton, and for magnetic black holes
in a background magnetic field.   The solutions are
(don't worry about how they were found!)
\eqn\ErnSol{\eqalign{ds^2& = \frac{\Lambda^2}{A^2(x-y)^2} \left[G(y)
dt^2 -
G^{-1}(y) dy^2 + G^{-1}(x) dx^2\right] +
\frac{G(x)}{\Lambda^2A^2(x-y)^2}d\phi^2\ ,\cr
A_\phi &= -\frac{2}{B\Lambda} \left[1+\half Bqx\right] + k\ .}}
Here
\eqn\seveneleven{G(\xi) = (1-\xi^2-r_+ A\xi^3)(1+r_- A\xi) \equiv -
r_+
r_- A^2\prod\limits^4_{i=1} (\xi - \xi_i)\ ,}
with ordered zeros
\eqn\seventwelve{\xi_1 = - \frac{1}{r_-A}\ , \ \xi_2, \xi_3, \xi_4\
,}
and
\eqn\seventhirteen{\Lambda\,(x,y) = (1+qBx)^2 + \frac{B^2
G(x)}{4A^2(x-y)^2}\ .}
Choice of $k$ corresponds to convention for location of Dirac string
singularities. $A,B,r_+,r_-$, and $q$ are parameters, to be thought
of
roughly as the acceleration, the magnetic field, the inner and outer
horizon radii, and the charge.  These identifications become exact in the
limit $qB\rightarrow 0$.  The latter three are related by
\eqn\sevenfourteen{q=\sqrt{r_+r_-}}
as for a free \RN\ black hole.  Furthermore, the charge, magnetic
field,
mass and acceleration should be related by an expression that simply
reduces to Newton's law $mA=qB$ at low accelerations,
$B\ll1$. Without this relation the solution has a string-like
singularity connecting the black holes.

\ifig{\Fig\ternst}{Shown is the lorentzian Ernst solution for an
accelerating mag\-net\-ic\-al\-ly-charged black hole.  The coordinates used in
(7.16) cover only the unshaded portion of the figure.
}{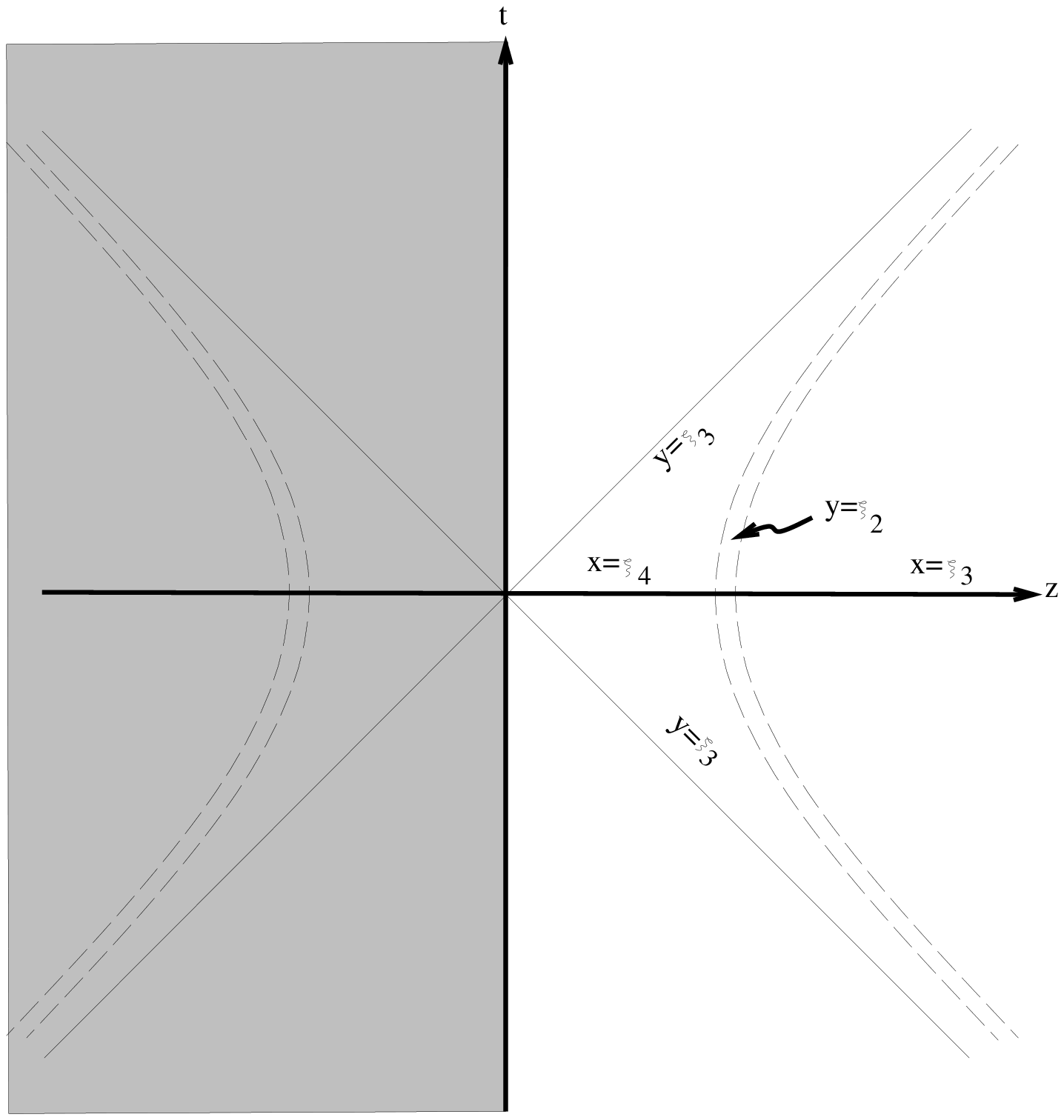}{3.50}

The solution \ErnSol\ appears rather complicated, but its structure can be
deduced by examining various limits.\foot{For more details, see \eg
\ \DGGH.} As $x\to y$ the solution asymptotes to
\eqn\sevenfifteen{\eqalign{ds^2 & =
\left(1+\frac{\tB^2\rho^2}{4}\right)^2
\left[-d{\hat t}^2 + dz^2 + d\rho^2\right] +
\left(1+\frac{\tB^2\rho^2}{4}\right)^{-2}
\rho^2 d{\hat \phi}^2\ ,\cr
A_{\hat \phi} & = \frac{\tB\rho^2}{2}\
\frac{1}{\left(1+\frac{\tB^2\rho^2}{4}\right)}\ ,}}
where $\tB = \tB(r_+,r_-,B)\simeq B$, and where a change
of coordinates has been
made.
This is the Melvin solution\refs{\Melv} --- due to the energy density
of the magnetic field, this is the closest one can come to
a uniform magnetic field in general relativity.  The solution \ErnSol\
naturally yields a
Rindler parametrization, and only covers half of the Melvin solution.
The rest is found by continuation. The region near the black hole
corresponds to $y\simeq \xi_2$, and in this limit the solution
asymptotes to the \RN\ solution.  The black hole follows a trajectory
with asymptotes $y=\xi_3$, corresponding to the acceleration horizon.
These features are illustrated in fig.~14.

Analytic continuation,
$t=i\tau$, gives a black hole moving on a circular trajectory, as in the
lower half of
fig.~13.  Regularity of the solution at the acceleration horizon requires a
specific periodicity for $\tau$ as in standard treatments of Rindler space.
However, regularity of the solution at the black hole horizon $y=\xi_2$
also requires periodic identification of $\tau$, in general with a
different period.
Demanding that these identifications match gives
another
condition on the parameters,
\eqn\pmatch{\xi_1-\xi_2-\xi_3+\xi_4 =0\ .}
This can be thought of as a condition
matching the Hawking temperature of the black hole and the
acceleration
temperature from its motion, which is necessary to find a
stationary solution with the black hole in thermodynamic equilibrium.

\ifig{\Fig\ttemp}{A sketch of the temperature vs. mass curve for a \RN\
black hole.}{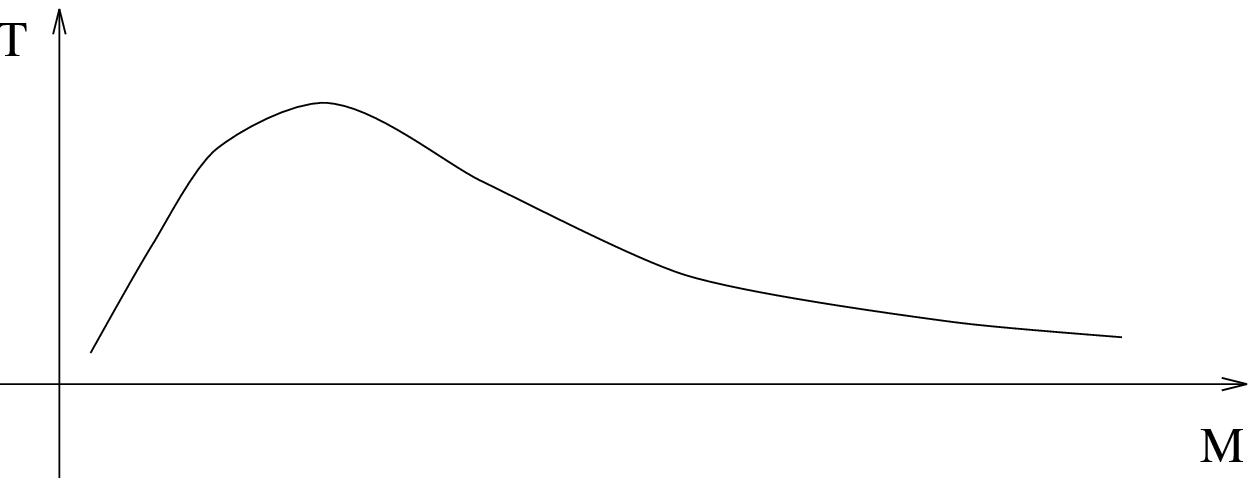}{1.4}

\ifig{\Fig\worms}{The spatial geometry of the time-symmetric
slice through the wormhole instanton.}{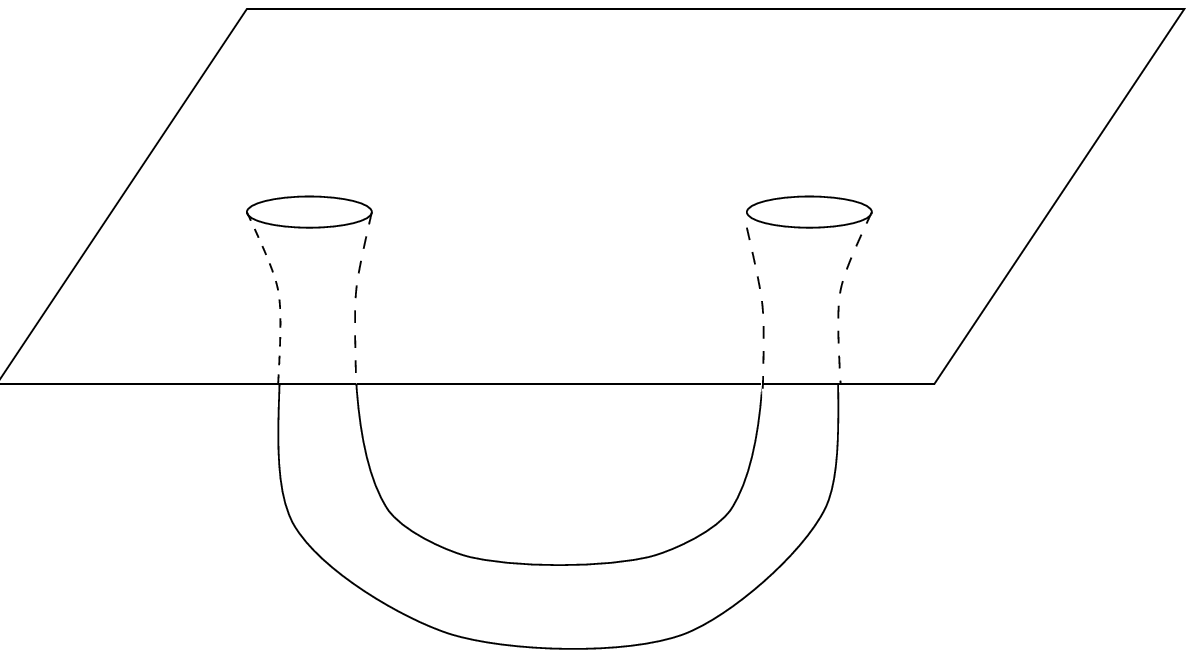}{1.5}

A sketch of the temperature vs. mass curve is in fig. 15.
The solution\refs{\GaSt,\DGKT}
to ${T_{BH}} = {T_{accel}}$ is given by taking
${M}$ slightly larger than $Q$. On the symmetric slice ${S}$ of
fig.~13, the three geometry is that of fig.~16
--near the black hole
it corresponds to a symmetric slice through the black hole horizon, as
in fig.~17. Thus it's geometry is that of a Wheeler wormhole with
ends of charge ${\pm Q}$. Another solution\refs{\Gibb,\DGGH}
takes advantage of the
ambiguity in the extremal limit ${M}{\rightarrow}{Q}$: here the
``throat'' near the horizon becomes infinitely long, and the periodic
identification is no longer fixed at the horizon. The geometry of the
symmetric slice is sketched in fig.~18.

\ifig{\Fig\rnpd}{A portion of the Penrose diagram for an $M>Q$ \RN\ black
hole.  Near the black hole, the geometry of fig.~16 is similar to that on
the slice $S$.}{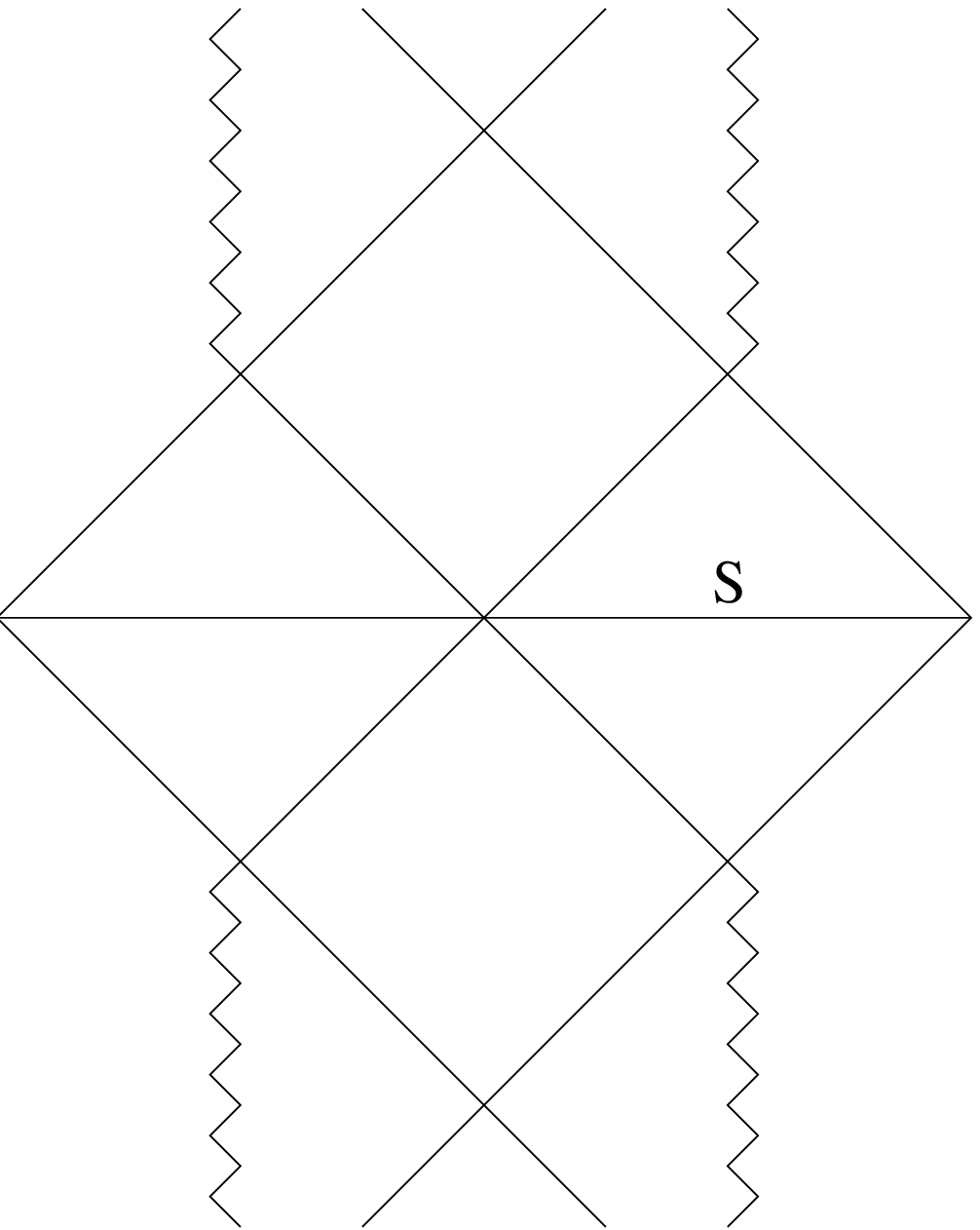}{2.0}

\ifig{\Fig\extsli}{The geometry of the symmetric
spatial slice through the extremal
instanton.}{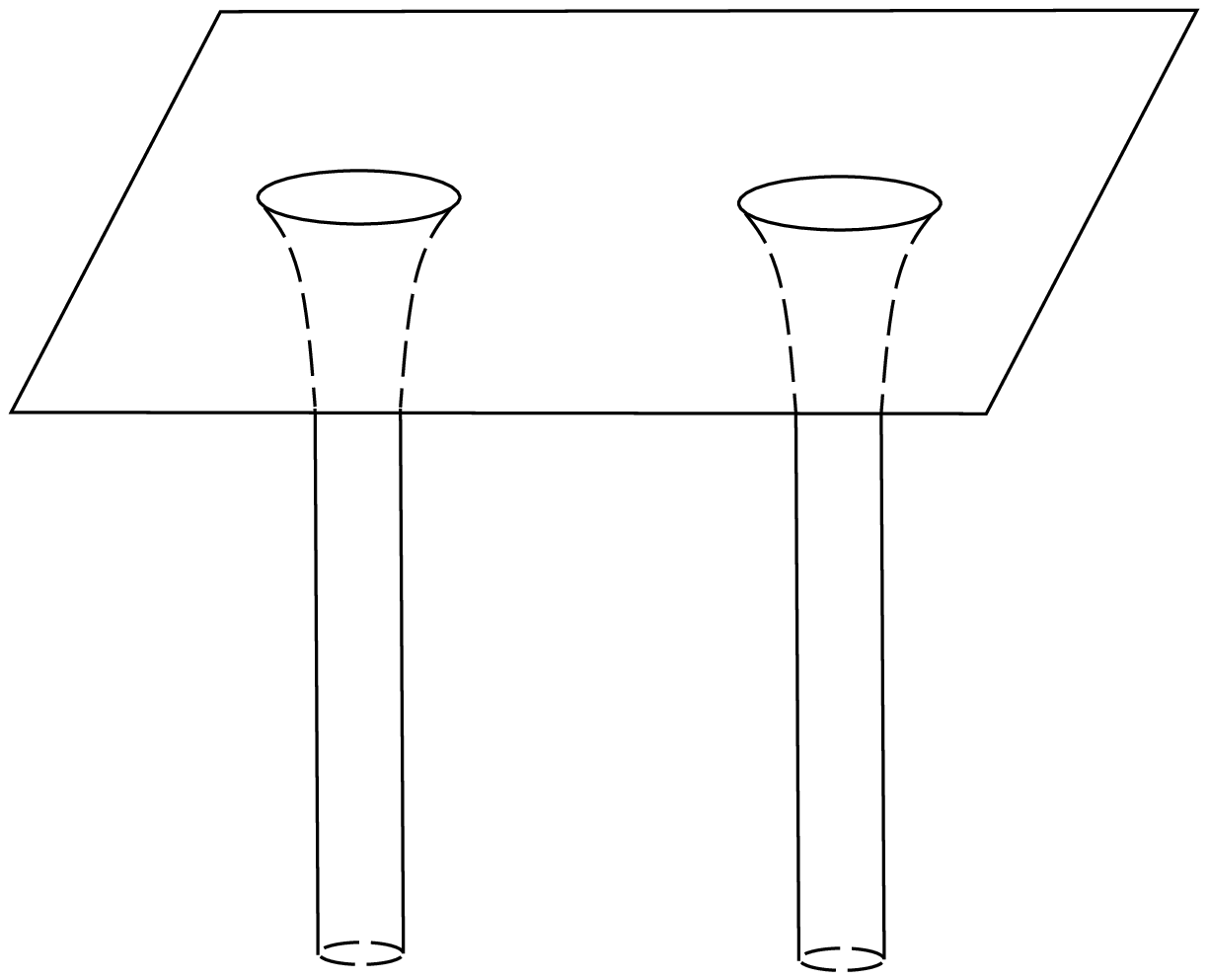}{2.0}

Necessary conditions for the validity of the semiclassical
expansion are ${q}\gg{1}$ (super-planckian black holes) and
${qB}\ll{1}$ (weak magnetic fields). To leading order in this
expansion the production rate is given by the instanton action,
\eqn\Apple{{\Gamma}\,\,{\propto}\,\,{e^{-S_{inst}}}\ .}
One can then show \refs{\GaSt,\GGS} that in terms of the physical charge
$Q$ and field $B$,
\eqn\Banana {{S_{inst}} = \frac{\pi Q}{B}\,\, ({1} + {\cal O}
({QB})),}
in agreement with \sevenseven.  However, as above one would
expect the rate to also be proportional to the number of black hole
internal states, which we have argued is infinite.

\ifig{\Fig\RNevap}{A portion of the Penrose diagram of an extremal \RN\
black hole, into which some matter has been dropped, and which subsequently
evaporates back to extremality.}{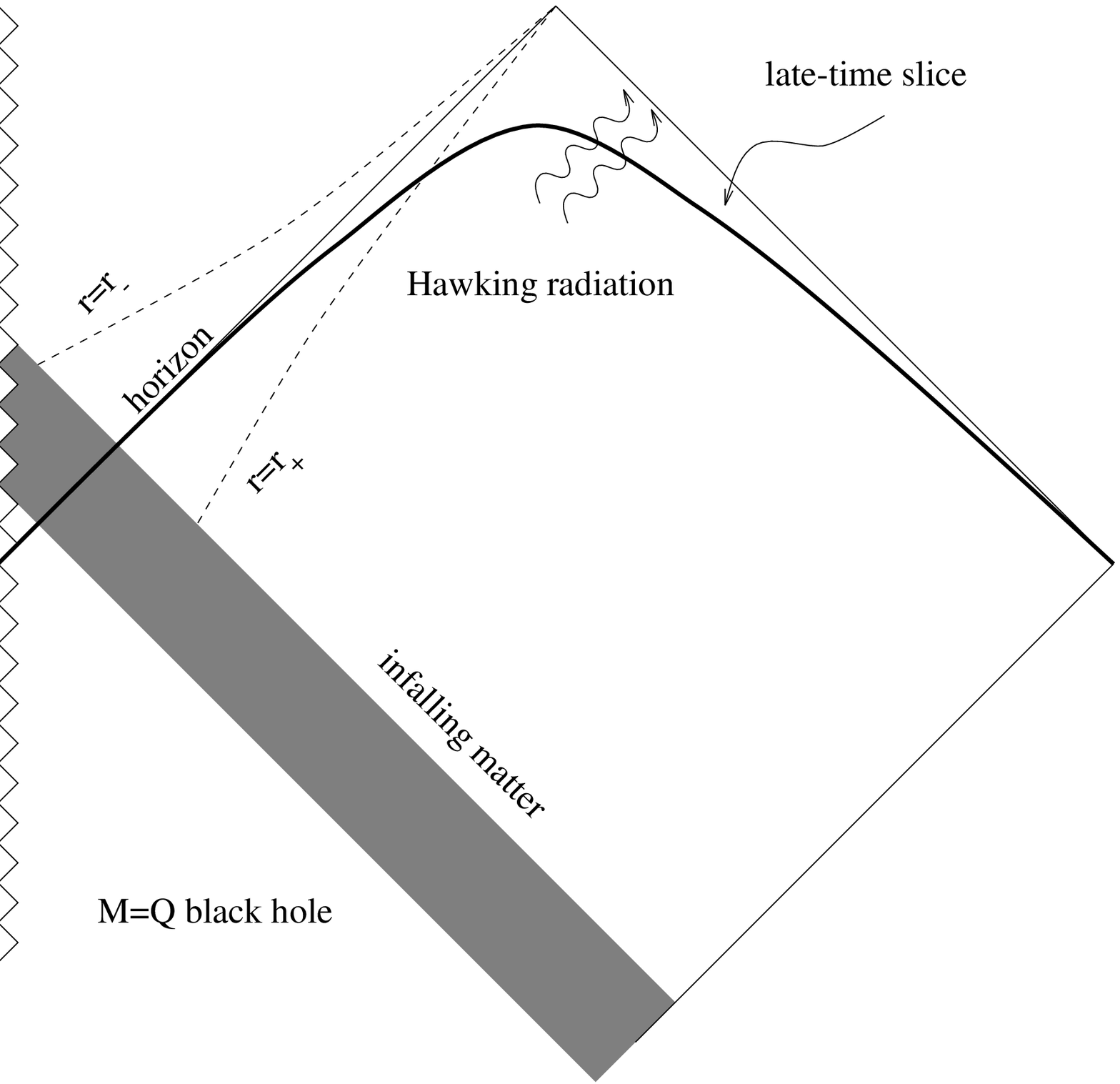}{2.8}

In order to see how the infinite states contribute, it's
useful to think about their description. In particular, suppose we
start with an extremal black hole, throw some matter into it, and let
it evaporate back to extremality\refs{\StTr}.
The Penrose diagram for this is
shown in fig. 19. In describing the evolution we are allowed to
choose any time slicing, and from our earlier discussion of the
instanton we expect a slicing with spatial slices that stay outside
the horizon to be a useful choice. At long times compared to the
evaporation time scale ${\sim}{Q^3}$, the mass excess from the
infalling matter will have been re-radiated in the form of Hawking
radiation, and this will asymptote to infinity. The state inside a
finite radius $R$ is therefore that of an extremal black hole until the
slice nears the incoming matter; see fig. 20. As
${t}{\rightarrow}{\infty}$, this matter is infinitely far down the
black hole throat, and is infinitely blueshifted. Clearly a
description of it using our time slicing requires planckian physics.

\ifigx{\Fig\rnsli}{A schematic picture of the state of the black
hole of fig.~19 as described on a late time slice that stays outside the
horizon.}{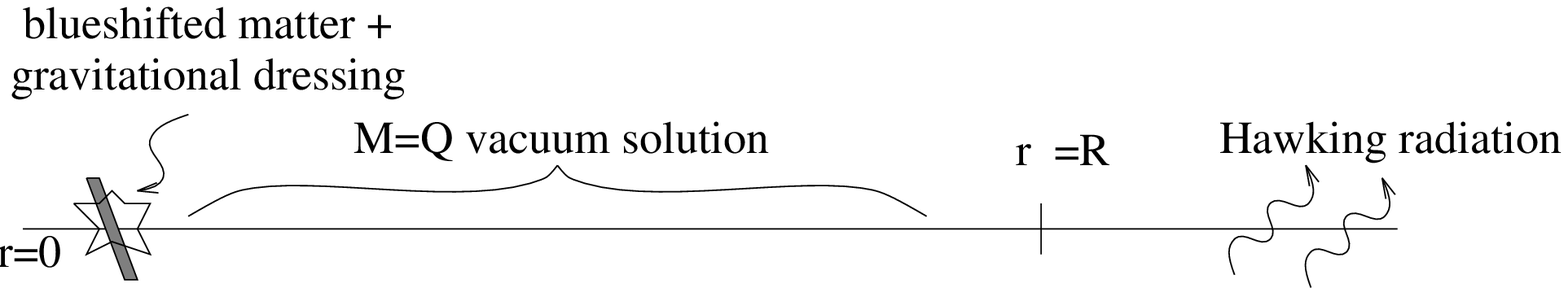}{4.5}

The contribution of such states to the pair-production rate
should come from the functional integral about the instanton, or, at
linear level, from the fluctuation determinant. The reason is that
the instanton only describes tunneling to the classical turning
point, but in computing the full tunneling rate we should consider
tunneling to the nearby configurations with gravitational and matter
perturbations, as described above. Tunneling to nearby states is via
paths near to the instanton, and it can be shown that
contributions of such paths gives the fluctuation determinant at the
linear level\refs{\BiCh}, or the full functional integral if interactions are
included.

The necessity for the outside observer to use Planck physics
in describing these states hints that perhaps Planck physics could
play an important role in computing the production rate\refs{\CILAR},
and in particular one might hope for it to be finite.

However, a careful examination of the instanton reveals that
for weak fields, ${QB}{<<}{1}$, the instanton closely approximates
the euclidean solution for a free black hole near the horizon. The
contribution to the functional integral from the vicinity of the
black hole should therefore be essentially the same as that to
${Tr}\,\,{e^{-\beta H}}$ from a free black hole, where the trace is
over black hole internal states, ${H}$ is the hamiltonian, and
${\beta}$ is given by the acceleration temperature. The appearance of
such a factor is in agreement with the effective field theory
computation, \proptopart. If the number of black hole internal
states is literally infinite, this factor would be expected to be infinite
as would be the production rate.

These issues are still being investigated \refs {\WABHIP},
and a definitive verdict on remnants is not in. Despite the apparent
need for a planckian description of these objects, the argument for
infinite production has endangered the viability of remnants. It may
be that black holes are only infinitely produced if they truly have a
finite number of internal states.

\newsec{Conclusions}

The quantum mechanics of black holes will no doubt remain a
fascinating open window on Planck scale physics for some
time to come. In particular, we can hope to sharpen our knowledge of
the properties of quantum gravity in our continued confrontation with
the information problem.   A diversity of opinions on its ultimate
resolution abounds, and this suggests the answer might be quite
interesting when finally discovered.

Black hole pair production should either tell us how to kill the remnant
scenario
or how to save it. In either case, it remains a rich and
interesting process, nontrivially combining the phenomena of Hawking
and acceleration radiation as well as other aspects of quantum
gravity.

The quantum mechanics of black holes has much left to teach
us.

\bigskip\bigskip\centerline{{\bf Acknowledgments}}\nobreak
I wish to thank the organizers for inviting me to Trieste to participate in
this most interesting school. I have benefited from discussions with T.
Banks, G. Horowitz, D. Lowe, W. Nelson, M.
Srednicki, A. Strominger, L. Susskind,
E. Verlinde, H. Verlinde, and many others.
This work was supported in part by DOE grant DOE-91ER40618 and
by NSF PYI grant PHY-9157463.

\listrefs

\end